\documentclass[10pt,a4paper,final]{iopart}

\usepackage{amsopn,amstext,amsbsy,setstack,amsthm,amssymb,eucal}
\usepackage[english]{babel}
\usepackage{mathrsfs}

\newcommand{\identity}{1\hspace{-0.231em}\mathrm{l}}        
\newcommand{\me}{\mathrm{e}}                                
\newcommand{\mi}{\mathrm{i}}                                
\newcommand{\difx}[1]{\mathrm{d}#1\;}                       
\newcommand{\difnx}[2]{\mathrm{d}^{#1}{#2}\;}               
\newcommand{\slashnabla}{\nabla\hspace{-0.75em}{/}\hspace{0.4em}} 
\DeclareMathOperator{\supp}{supp}                           
\newcommand{\fathat}{\widehat{\phantom{\big]}}}             
\newcommand{\extd}{\mathrm{d}}                              
\newcommand{\norder}[1]{{:}#1{:}}                           
\newcommand{\transp}[1]{{#1}^{\mathrm{T}}}                  

\newcommand{\Wtwo}{W}
\newcommand{\Wzerotwo}{W_0}
\newcommand{\WtwoAB}{W_{AB}}

\newcommand{\Ws}{\mathsf{W}}

\DeclareMathOperator{\Spin}{Spin}                               
\DeclareMathOperator{\SL}{SL}                                   
\DeclareMathOperator{\SO}{SO}                                   

\DeclareMathOperator{\diag}{diag}

\DeclareMathOperator{\WF}{WF}                                   

\newcommand{\LL}{\mathscr{L}}                                   
\newcommand{\LLpo}{\LL^{\uparrow}_+}                            

\newcommand{\Mc}{\mathcal{M}}                                   
\newcommand{\Nc}{\mathcal{N}}
\newcommand{\Qc}{\mathcal{Q}}

\newcommand{\Mb}{{\boldsymbol{\Mc}}}
\newcommand{\Nb}{{\boldsymbol{\Nc}}}

\newcommand{\gb}{\mathbf{g}}                                    
\newcommand{\gammb}{\boldsymbol{\gamma}}                        

\newcommand{\RR}{\mathbb{R}}                                    
\newcommand{\CC}{\mathbb{C}}                                    

\newcommand{\DD}{\mathscr{D}}                                   
\newcommand{\DDsp}{\DD_{\text{sp}}}                             
\newcommand{\DDcosp}{\DD_{\text{cosp}}}                         
\newcommand{\DDd}{\DD_{\text{double}}}                          
\newcommand{\DDdM}{\DD_{\text{double},\Mb}}
\newcommand{\DDdMp}{\DD_{\text{double},\Mb'}}

\newcommand{\Ff}{\mathfrak{F}}                                  
\newcommand{\Tf}{\mathfrak{T}}

\newcommand{\Ssp}{S_{\text{sp}}}                                
\newcommand{\Scosp}{S_{\text{cosp}}}                            

\newcommand{\Cinfzero}{C^\infty_0}                              
\newcommand{\Ctinfzero}{\widetilde{C^\infty_0}}

\newcommand{\bts}{\beta_{t\ast}}
\newcommand{\bhts}{\widehat{\beta}_{t\ast}}
\newcommand{\btts}{\widetilde{\beta}_{t\ast}}

\newcommand{\id}{\text{id}}
\newcommand{\kb}{{\boldsymbol{k}}}

\newcommand{\Thetab}{{\boldsymbol{\Theta}}}

\newcommand{\fs}{{\sf f}}
\newcommand{\ts}{{\sf t}}
\newcommand{\Ts}{{\sf T}}

\newtheorem{theorem}{Theorem}

\newtheorem{lemma}[theorem]{Lemma}

\newtheorem{proposition}[theorem]{Proposition}

\begin{document}

\title[An explicit QWEI for Dirac fields in curved spacetimes]
  {An explicit quantum weak energy inequality for Dirac fields in curved spacetimes}
\author{S P Dawson and C J Fewster}
\address{Department of Mathematics, University of York, Heslington, York YO10 5DD, UK}
\eads{\mailto{spdawson@gmail.com}, \mailto{cjf3@york.ac.uk}}

\begin{abstract}
The quantized Dirac field is known, by a result of Fewster and Verch, to satisfy a Quantum Weak Energy Inequality (QWEI)
on its averaged energy density along time-like curves in arbitrary four-dimensional globally hyperbolic spacetimes.
However, this result does not provide an explicit form for the bound. By adapting ideas from the earlier work, we give a
simplified derivation of a QWEI for the Dirac field leading to an explicit bound. The bound simplifies further in the
case of static curves in static spacetimes, and, in particular, coincides with a result of Fewster and Mistry in
four-dimensional Minkowski spacetime. We also show that our QWEI is compatible with local covariance and derive a simple
consequence.
\end{abstract}


\section{Introduction}

No quantum field (obeying Wightman axioms) can have a nontrivial energy density whose expectation values are always
nonnegative \cite{Epstein/Glaser/Jaffe}. Moreover, in all models studied (and all models with certain scaling behaviour
\cite{Fewster_review}) the energy density at any given point can be made arbitrarily negative by a suitable choice of
the state of the field. Taken at face value, these surprising facts would raise concerns about the possibility of
violations of the second law of thermodynamics \cite{Ford 1978} or other instabilities arising from extended
distributions of `exotic matter'. However, as was originally realized by Ford \cite{Ford 1978}, quantum field theory
appears to contain mechanisms---ultimately related to the uncertainty principle---which constrain the magnitude and
duration of violations of the classical energy conditions. These are expressed by Quantum Energy Inequalities (QEIs),
which give lower bounds on the averages of the stress-energy tensor taken along the world-line of an observer, or over a
spacetime volume. When the energy density itself is averaged, the more specific term Quantum Weak Energy Inequality
(QWEI) is often used.

QEIs are known for scalar, spin-$\frac{1}{2}$ and spin-$1$ free fields in arbitrary globally hyperbolic spacetimes, and
also for all positive energy unitary conformal field theories in two-dimensional Minkowski spacetime
\cite{Fewster/Hollands}. Reviews of the literature on QEIs and their applications (and related issues) may be found in
\cite{Fewster_review,Roman_review}.

The general results in arbitrary spacetimes are all obtained using techniques drawn from microlocal analysis. However,
while the QWEIs obtained for the scalar \cite{Fewster 2000} and spin-$1$ \cite{Fewster/Pfenning_1} fields follow a
common pattern and lead to explicit bounds, the general spin-$\frac{1}{2}$ QWEI obtained by Fewster and Verch
\cite{Fewster/Verch_1} was proved by breaking the averaged energy density into a number of terms which were bounded by
undetermined (although finite) constants. The only explicit Dirac QWEI known in four dimensions was obtained by Fewster
and Mistry \cite{Fewster/Mistry} in Minkowski spacetime, based on an identity discovered in \cite{Fewster/Verch_1}. The
same approach may be applied to Rarita--Schwinger fields in Minkowski spacetime \cite{Yu/Wu,Hu/Ling/Zhang}, and has been
adapted to the case of two-dimensional Minkowski spacetime by Dawson \cite{Dawson}; different methods have been used to
treat the specific case of massless Dirac fields in two dimensional curved spacetimes \cite{Vollick}.

In this paper, we will combine the language and formalism of \cite{Fewster/Verch_1} with the general approach of
\cite{Fewster/Mistry}; this tactic results in a bound that is easily compared and contrasted with the
other general world-line QWEIs \cite{Fewster 2000, Fewster/Pfenning_1}.

Our result may be stated as follows. Let $(\Mc,\gb)$ be a four-dimensional globally hyperbolic Lorentzian manifold,
which is orientable and time-orientable, and suppose that such orientations have been chosen, along with a spin
structure. As we review in \S\ref{Section: Dirac field on a curved spacetime}, the Dirac field may be formulated on
$(\Mc,\gb)$, along with an appropriate notion of Hadamard states, for which the stress tensor may be defined by
point-splitting. Let $\omega_0$ be any Hadamard state of this theory and use it as a reference state to define the
normal-ordered stress tensor $\norder{T_{\mu\nu}}$ as described in \cite{Fewster/Verch_1}. If $\gamma:I\rightarrow\Mc$ is a
proper-time parameterization of a smooth future-directed time-like curve in $(\Mc,\gb)$ for some open interval
$I\subseteq\RR$, then the normal-ordered energy density, as seen by an observer with world-line $\gamma$, is
\begin{equation}\label{definition of rho}
  \langle\norder{\rho}\rangle_\omega\left(\tau\right) :=
    \omega\left(u^\mu\left(\tau\right)u^\nu\left(\tau\right)\norder{T_{\mu\nu}}\left(\gamma\left(\tau\right)\right)\right),
\end{equation}
where $u(\tau)=\dot{\gamma}(\tau)$ is the unit tangent vector to $\gamma$ at
$\tau$. We will make some comments on the nature of the normal ordered
stress tensor at the end of this section.

Using a combination of parallel transport and Fermi--Walker
transport\footnote{See \ref{Appendix: Fermi-Walker_transport}, for a
review.}, we construct a
local section $E$ of the spin bundle near $\gamma$ such that the induced tetrad $e=(e_a)$ $(a=0,1,2,3)$ obeys
$e_0(\gamma(\tau))=u(\tau)$. Together with the reference state, this
permits us to define distributions near $\gamma$ by
\begin{eqnarray}
  \Ws_0\left(f,h\right)&:=&\delta^{AB}\omega_0\left(\Psi^{+}\left(fE_A\right)\Psi\left(hE_B^+\right)\right) \\
  \Ws_0^\Gamma\left(f,h\right)&:=&\delta^{AB}\omega_0\left(\Psi\left(hE_B^+\right)\Psi^{+}\left(fE_A\right)\right),
\end{eqnarray}
where $E_A$ ($A=1,2,3,4$) form a spin-frame induced by $E$, and
$\Psi$ and $\Psi^+$ are the Dirac field and its Dirac adjoint. These
distributions may be shown to be independent of the freedom
in the construction of $E$.

With these assumptions, our main result is the following:
\begin{theorem}\label{thm:main}
  For any real-valued weight $g\in \Cinfzero(I)$, and any Hadamard state $\omega$ of the Dirac field on $(\Mc,\gb)$,
  \begin{equation}\label{eq: main result}
    \int\difx{\tau}\langle\norder{\rho}\rangle_\omega\left(\tau\right)g\left(\tau\right)^2
      \geq
    -\frac{1}{2\pi}\int_0^{\infty}\difx{\mu}\mu \left( S_\mu +S^\Gamma_\mu \right) > -\infty,
  \end{equation}
  where $S_\mu$ and $S^\Gamma_\mu$ are positive functions, decaying rapidly as $\mu\rightarrow+\infty$, and depending
  on $\gamma$, $g$, and the distributions $\Ws_0$ and $\Ws_0^\Gamma$.
  They are defined by
  \begin{eqnarray}\label{eq: defn of S mu}
    S_\mu &:=& \left[g\otimes g\gamma_2^\ast \Ws_0\right]\fathat\left(-\mu,\mu\right), \\
    S^\Gamma_\mu &:=& \left[g\otimes g\gamma_2^\ast \Ws_0^\Gamma\right]\fathat\left(\mu,-\mu\right),
  \end{eqnarray}
  where $\gamma_2^\ast$ denotes the pull-back by
  $\gamma_2(\tau,\tau^\prime):=(\gamma(\tau),\gamma(\tau^\prime))$ and the
  hat denotes the Fourier transform (according to conventions given below).
\end{theorem}

It is worth making a few remarks before we proceed. First, the bound of equation (\ref{eq: main result}) is far more explicit than that
given in \cite{Fewster/Verch_1}. Second, $\Ws_0$ may be regarded as a point-split unrenormalized charge density for the reference state,
and $\Ws_0^\Gamma$ is closely related to the corresponding quantity for the charge conjugate state. If the reference state $\omega_0$ is
charge-conjugation invariant, this relation causes the bound to simplify by virtue of the relation $S_\mu^\Gamma=S_\mu$.

Third, the bound also takes on a simpler form for a static observer in a static spacetime, as will be discussed in
\S\ref{sec:static}. In particular, the Minkowski spacetime result of \cite{Fewster/Mistry} is recovered as a special
case in \S\ref{subsec:Minkowski}. Fourth, as we will show in \S\ref{Section:covariance}, our QWEI is a locally covariant
difference QWEI, in the sense recently developed in \cite{Fewster/Pfenning_2}. Fifth: although explicit, the bound is
not expected to be optimal. As already mentioned, the general technique of \cite{Fewster/Mistry} was applied to the case
of two-dimensional flat spacetime in \cite{Dawson}; in the massless limit, the result was weaker than the optimal bounds
of Vollick \cite{Vollick} (cf.\ also \cite{Flanagan,Flanagan 2002}).

Finally, we reiterate that our bound is a \emph{difference} QWEI; that is, it applies to the normal ordered energy density with
respect to a reference state $\omega_0$, rather than the renormalized stress tensor $T_{\mu\nu}^{\text{ren}}$, which is covariantly
defined without the use of a reference state and exhibits a trace anomaly. Now, it is one of the Wald axioms \cite{Wald_qft} for
stress tensor renormalization that
$\omega(T_{\mu\nu}^{\text{ren}}(x))-\omega_0(T_{\mu\nu}^{\text{ren}}(x))=\omega(\norder{T_{\mu\nu}}(x))$, so we may replace
$\norder{\rho}$ by $\rho^{\text{ren}}$ in our result (\ref{eq: main result}) at the expense of adding
$\int_\gamma\difx{\tau}\langle\rho^{\text{ren}}(\gamma(\tau))\rangle_{\omega_0}g(\tau)^2$ to the right-hand side. Since $\omega_0$ is
Hadamard, this is a finite modification and the renormalized energy density is also seen to be bounded from below (thus constituting
an {\em absolute} QWEI). In the scalar case, an absolute QWEI may be obtained without appealing to a reference state
\cite{Fewster/Smith}, and one expects that this can also be done for Dirac fields. It is also worth noting that Wald's prescription
for stress tensor renormalization \cite{Wald1978} involves the addition of certain terms `by hand' to ensure that the expectation
value of the stress tensor is conserved, and vanishes in the Minkowski vacuum state. In the scalar case, it has recently been shown by
Moretti \cite{Moretti} that an ingenious modification of the stress tensor, which leaves the classical expression unchanged for
solutions to the Klein--Gordon equation, removes the necessity for such additions and gives the same final result. It would be
interesting to see whether a similar programme can be carried out for the free Dirac field (cf. e.g., \cite{Fujikawa}); of course, our QWEI would still apply, as the expectation values are unchanged.

\subsection{Definitions and notational conventions}

We work in `natural' units, so that $\hbar = c = 1$. Lower (respectively, upper) case Latin characters will be used to
label tetrad (respectively, spinor) indices. Tetrad indices take values $0$--$3$; spinor indices take values $1$--$4$.
Spacetime indices are indicated by lower-case Greek characters.

The smooth map $\me_k:\RR^n\rightarrow\CC$ is defined by
\begin{equation}
  x\mapsto\me_k\left(x\right):=\me^{\mi k\cdot x},
    \quad k\in\RR^n.
\end{equation}
We define the Fourier transform $\widehat{f}$ of a function $f\in L^1(\RR^n)$ using the `usual non-standard' convention
\begin{equation}
  \widehat{f}\left(k\right):=\int\difnx{n}{x} f\left(x\right)
    \me_k\left(x\right).
\end{equation}
In terms of these definitions, the Fourier transform of a distribution $u\in\mathscr{E}^\prime(\RR^n)$ is defined by
\begin{equation}
  \widehat{u}\left(k\right):=u\left(\me_k\right).
\end{equation}

\section{The quantized Dirac field on a curved spacetime}\label{Section: Dirac field on a curved spacetime}

\subsection{Geometry of Dirac fields on curved spacetimes} \label{subsec:Geometry}

In order for the present work to be reasonably self-contained, we will summarize here the essential features  that allow
one to formulate a meaningful description of a Dirac field on a spacetime manifold. We shall adhere closely to the
definitions, terminology and notation of \cite{Fewster/Verch_1,Dimock}, and much of the following material is drawn
directly from those sources.

We begin by defining $\LL$ to be the group of $4\times 4$ real Lorentz matrices $\Lambda^{a}_{\phantom{a}b}$ with the defining property
that $\eta_{ab}\Lambda^{a}_{\phantom{a}{c}}\Lambda^{b}_{\phantom{a}{d}}=\eta_{cd}$, where $\eta=\diag(+1,-1,-1,-1)$ is the usual
Minkowski metric. The identity connected component of $\LL$ is the subgroup of proper orthochronous Lorentz matrices $\LLpo$. We also
fix a set of $4\times 4$ Dirac matrices $\gamma_a$ $(a=1,\ldots,4)$ obeying
\begin{equation}\label{eq:Dirac_def}
  \gamma_a \gamma_b+\gamma_b\gamma_a = 2\eta_{ab}\identity
\end{equation}
and belonging to a standard representation, which means that
\begin{equation}
  \gamma_0^\dag = \gamma_0 \quad\text{and}\quad \gamma_k^\dag = -\gamma_k
\end{equation}
for $k=1,2,3$, where ${}^\dag$ denotes the usual Hermitian transpose on matrices.

According to a general theorem of Pauli \cite{Pauli}, any two sets of Dirac matrices are intertwined by a nonsingular matrix, unique up
to a scalar multiple. In particular, there is a nonsingular matrix $C$ such that
\begin{equation}\label{eq:C_def}
  C\gamma_a = -\transp{\gamma_a} C.
\end{equation}
The matrix $C$ may be shown to be antisymmetric; moreover, in a standard representation $C$ may be normalized so that
\begin{equation}
  C^\dag C = - C^2 = \identity
\end{equation}
which fixes it up to an overall sign. In consequence we also have
\begin{equation}
  C = \overline{C} = -C^\dag.
\end{equation}

The \emph{spin group} $\Spin(1,3)$ is the group of matrices $S\in\SL(4,\CC)$ such that
\begin{equation}\label{eq:S_def}
  S\gamma_a S^{-1} = \gamma_b \Lambda^{b}_{\phantom{b}a}
\end{equation}
for some coefficients $\Lambda^{b}_{\phantom{b}a}$ which, by equation (\ref{eq:Dirac_def}), are necessarily the components of a Lorentz
matrix. It can be shown that the map $S\mapsto\Lambda(S)$ is a two-to-one covering homomorphism from the identity-connected component
$\Spin_0(1,3)$ to $\LLpo$, with kernel $\{\identity,-\identity\}$, so we also have $\Spin_0(1,3)\cong \SL(2,\CC)$.

By considering the transpose of (\ref{eq:S_def}) and using the definition of $C$ it is easy to show that
$CS\gamma_a(CS)^{-1}=\transp{S^{-1}}C\gamma_a (\transp{S^{-1}}C)^{-1}$ for each $a$. By Pauli's theorem this entails that $CS = k_S
\transp{S^{-1}}C$ for each $S$, where the constants $k_S$ are readily seen to obey $k_S^4=1$ on considering determinants. Using
continuity and $k_{\identity}=1$, we may conclude that $k_S=1$ for all $S$, yielding
\begin{equation}\label{eq:S_C}
  \transp{S}C = CS^{-1}
\end{equation}
for all $S\in\Spin_0(1,3)$. In a similar way, we may use the identity
\begin{equation}
  \gamma_0\gamma_a = \gamma_a^\dag\gamma_0,
\end{equation}
which holds in standard representations, to deduce that
\begin{equation}\label{eq:S_gamma0}
  S^\dag \gamma_0 = \gamma_0 S^{-1}
\end{equation}
for all $S\in\Spin_0(1,3)$. This has a useful consequence: if $\Lambda(S)$ is a pure rotation, so that $S\gamma_0S^{-1}=\gamma_0$, then
$S^\dag\gamma_0=\gamma_0 S^{-1}=S^{-1}\gamma_0$, and hence $S$ is unitary.

Turning to the curved spacetime setting, we assume that an orientation and time orientation have been chosen on $(\Mc,\gb)$. The frame
bundle $F(\Mc,\gb)$ is the bundle of oriented and time-oriented orthonormal frames $e=(e_a)_{a=0,\ldots,3}$ over $(\Mc,\gb)$ with the
convention that $e_0$ is time-like and future pointing. This is a principal $\LLpo$-bundle, with the right action
\begin{equation}
  \left(R_\Lambda e\right)_a = e_b\Lambda^b_{\phantom{b}a}.
\end{equation}
A \emph{spin structure} on $(\Mc,\gb)$ is a principal $\Spin_0(1,3)$-bundle $S(\Mc,\gb)$ over $(\Mc,\gb)$ together with
a fibre-bundle homomorphism $\psi:S(\Mc,\gb)\rightarrow F(\Mc,\gb)$ such that $\psi$ intertwines the right action of the
structure groups on these bundles:
\begin{equation}
  \psi\circ R_S = R_{\Lambda\left(S\right)}\circ \psi.
\end{equation}
Spin structures necessarily exist on the spacetimes we consider, but are not necessarily unique. We assume that a particular spin
structure has been chosen from now on.

Spinor fields are now defined as sections of another bundle $D\Mc$, which is an associated $\Spin_0(1,3)$-bundle
\begin{equation}
  D\Mc = S\left(\Mc,\gb\right)\ltimes_{\Spin_0\left(1,3\right)}\CC^4.
\end{equation}
That is, the fibre of $D\Mc$ at $p\in\Mc$ consists of equivalence classes $[T,x]_p$ for $T\in S(\Mc,\gb)_p$, $x\in\CC^4$
(considered as a $4$-dimensional complex column vector) where $[T^\prime,x^\prime]_p=[T,x]_p$ if and only if
$T^\prime=R_S^{-1}T$, $x^\prime=Sx$ for some $S\in\Spin_0(1,3)$. The upshot is that $D\Mc$ has fibre $\CC^4$ at each
point $p\in\Mc$ and a left action of $\Spin_0(1,3)$ given by
\begin{equation}
  L_S\left[T,x\right]_p = \left[T,Sx\right]_p.
\end{equation}
The dual bundle $D^\ast\Mc$ may be constructed similarly, with fibres consisting of equivalence classes
$[T,\ell]_p^\ast$ for $T\in S(\Mc,\gb)_p$ and $\ell\in\CC^4$ (considered as a $4$-dimensional complex row vectors) with
$[T^\prime,\ell^\prime]_p^\ast=[T,\ell]_p^\ast$ if and only if $T^\prime=R_S^{-1}T$, $\ell^\prime=\ell S^{-1}$ for some
$S\in\Spin_0(1,3)$. Elements of $D\Mc$ are called spinors, while elements of $D^\ast\Mc$ are called cospinors, and have
a natural dual action on spinors: if $v_p=[T,\ell]^\ast_p$ and $u_p=[T,x]_p$, then
\begin{equation}
  v\left(u\right)\mid_p = \ell\cdot x
\end{equation}
where the dot denotes the usual matrix multiplication.

We may now define the Dirac adjoint and charge conjugation maps. The Dirac adjoint $u\mapsto u^{+}$ maps antilinearly
between spinors and cospinors so that
\begin{equation}
  \left(\left[T,x\right]_p\right)^{+} = \left[T,x^\dag\gamma_0\right]_p^\ast,
\end{equation}
which is well-defined owing to equation (\ref{eq:S_gamma0}). The inverse map is
also denoted in the same way. Charge conjugation $u\mapsto u^c$ is an
antilinear map of $D\Mc$ to itself defined by
\begin{equation}
  \left(\left[T,x\right]_p\right)^c = \left[T,\gamma_0 C^\dag\overline{x}\right]_p
\end{equation}
which is well-defined owing to equations (\ref{eq:S_gamma0}) and (\ref{eq:S_C}). The definition is extended to
$D^\ast\Mc$ by duality:
\begin{equation}
  \left(\left[T,\ell\right]_p^\ast\right)^c = \left[T,\overline{\ell}C\gamma_0\right]^\ast_p
\end{equation}
which entails that $v^c(u^c)=\overline{v(u)}$. Both spinors and cospinors obey the identities $\psi^{cc}= \psi$,
$\psi^{c+}=-\psi^{+c}$.

If $B$ is any bundle over $\Mc$, we use the notation $C^\infty(B)$ to denote the space of smooth sections of $B$, and
$\Cinfzero(B)$ for those of compact support. In particular, we will denote $\DDsp=\Cinfzero(D\Mc)$ and
$\DDcosp=\Cinfzero(D^\ast\Mc)$, endowed with their usual topologies, for spaces of smooth compactly supported (co)spinor test fields, as in
\cite{Fewster/Verch_1}.

Finally, any (local) section $E$ of $S(\Mc,\gb)$ determines a (local) frame field $e = (e_0,\ldots,e_3) = \psi\circ E$
and (local) sections $E_A$ of $D\Mc$ such that $E_A(p)=[E_p,b_A]_p$, where $b_A$ $(A=1,\ldots,4)$ is the standard basis
in $\CC^4$. These induce a dual frame $(e^0,\ldots,e^3)$ and dual sections $E^A$ of $D^\ast\Mc$ by $e^a(e_b) =
\delta^a_b$ and $E^A(E_B)=\delta^A_B$, and permit arbitrary mixed tensor-spinor fields to be expressed in component
form. In particular, $\gammb\in C^\infty(T^\ast\Mc \otimes D\Mc \otimes D^\ast\Mc)$ is defined to have
components $\gamma_{a\phantom{A}B}^{\phantom{a}A}$ ($=$ the matrix components of the Dirac matrix $\gamma_a$) in some
(and hence any) such system.

\subsection{The Dirac equation}

As usual, the metric induces a covariant derivative $\nabla$ on $C^\infty(T\Mc)$; if $k=k^b e_b\in C^\infty(T\Mc)$ is a
smooth section, then we have
\begin{equation}\label{defn_of_C_symbols}
  \nabla k=\left(\nabla_g k^c\right) e^g\otimes e_c=\left(\partial_b k^a+\Gamma^a_{bf}k^f\right)e^b\otimes e_a
\end{equation}
where the second equation defines the Christoffel connection coefficients $\Gamma^a_{bd}$.

In turn, a further covariant derivative, which we also denote by $\nabla$, is induced on $C^\infty(D\Mc)$. If
$(e_0,\ldots,e_3)$ and $(E_A)_{A=1}^3$ are induced by a section $E$ in $S(\Mc,\gb)$ and $f=f^A E_A$ is a local section
in $D\Mc$, then $\nabla f\in C^\infty(T^\ast\Mc \otimes D\Mc)$ is given by
\begin{equation}
  \nabla f = \left(\nabla_b f^A\right)e^b\otimes E_A =
  \left(\partial_b f^A + \sigma_{b\phantom{A}B}^{\phantom{b}A}f^B\right)e^b\otimes E_A.
\end{equation}
The connection coefficients $\sigma_{b\phantom{A}B}^{\phantom{b}A}$ appearing in the second equation are defined by
\begin{equation}
  \sigma_{b\phantom{A}B}^{\phantom{b}A} = -\frac{1}{4}\Gamma^a_{bd}\gamma_{a\phantom{A}C}^{\phantom{a}A}
    \gamma^{dC}_{\phantom{dC}B},
\end{equation}
and
\begin{equation}
  \partial_b f^A = \extd f^A\left(e_b\right),
\end{equation}
where $\extd f^A$ is the exterior derivative of the function $f^A$.

The action of $\nabla$ can be extended uniquely to cospinor and mixed spinor-tensor fields by imposing the usual
requirements that the covariant derivative be Leibniz and that it commute with arbitrary contractions. Thus, for
example, if $h=h_B E^B$ is a cospinor field, then the components of $\nabla h=\nabla_b h_B e^b\otimes E^B$ are
\begin{equation}
  \nabla_b h_B = \partial_b h_B - h_C\sigma_{b\phantom{C}B}^{\phantom{b}C}.
\end{equation}
With the covariant derivative defined in this way, it follows that $\nabla\gammb=0$.

The first-order differential equation
\begin{equation}
  \left(-\mi\slashnabla+m\right)u = 0
\end{equation}
for the spinor field $u\in C^\infty(D\Mc)$ is known as the \emph{Dirac equation}. The corresponding Dirac equation for
the cospinor field $v\in C^\infty(D^\ast\Mc)$ is
\begin{equation}
  \left(\mi\slashnabla+m\right)v = 0.
\end{equation}
The constant $m\geq0$ is interpreted as the field mass. As usual, $\slashnabla$ is the \emph{Dirac operator}, and maps
(co)spinor field to (co)spinor fields by
\begin{eqnarray}
  \slashnabla f = \left(\slashnabla f\right)^A E_A = \eta^{ab}\gamma_a{}^A{}_B\left(\nabla_b f^B\right) E_A, \\
  \slashnabla h = \left(\slashnabla h\right)_B E^B = \eta^{ab}\left(\nabla_b h_C\right) \gamma_a{}^C{}_B E^B,
\end{eqnarray}
where $f=f^A E_A\in C^\infty(D\Mc)$ and $h = h_B E^B\in C^\infty(D^\ast\Mc)$.

The advanced ($-$) and retarded ($+$) fundamental solutions, in the spinor case, are continuous linear maps
\begin{equation}
  \Ssp^{\pm}:\Cinfzero\left(D\Mc\right)\rightarrow C^\infty\left(D\Mc\right)
\end{equation}
such that
\begin{equation}
  \left(-\mi\slashnabla+m\right)\Ssp^{\pm} u = u = \Ssp^{\pm}\left(-\mi\slashnabla+m\right)u,
\end{equation}
and so that $\supp \Ssp^{\pm}u\subset J^{\pm}(\supp{u})$, where $J^{\pm}(\supp{u})$ is the causal future($+$)/causal
past($-$) of $\supp u$. The fundamental cospinor solutions $\Scosp^{\pm}$ are similarly
defined, and the
retarded-minus-advanced fundamental solutions are then written as
\begin{equation}
  \Ssp = \Ssp^+ - \Ssp^-\quad\text{and}\quad \Scosp = \Scosp^+ - \Scosp^-.
\end{equation}

\subsection{The field algebra}

Let $\DDd=\DDcosp \oplus \DDsp$ be the space of `doubled' test (co)spinors, on which we may define operators
\begin{equation}
  D:=\mi\left(\begin{array}{cc} \mi\slashnabla+m  &   0 \\
                                    0             &   -\mi\slashnabla+m \end{array}\right),\quad
  S:=\mi\left(\begin{array}{cc} \Scosp            &   0 \\
                                    0             &   \Ssp \end{array}\right)
\end{equation}
($D=D_{\rhd}$ and $S=S_{\lhd}$ in the notation of \cite{Fewster/Verch_1}) and an antilinear map
\begin{equation}
  \Gamma\left(\left[\begin{array}{c} h \\ f \end{array}\right]\right) = \left[\begin{array}{c} f^+ \\ h^+ \end{array}\right].
\end{equation}
We use $\DDd$ to label a set of abstract objects: to each $F\in\DDd$, we associate an object $\Xi(F)$. We may now define the \emph{field
algebra} to be a unital $\ast$-algebra $\Ff(\Mc,\gb)$ consisting of all (finite) polynomials in the $\Xi(F)$, their adjoints
$\Xi(F)^\ast$, and the identity $\identity$, subject to the following relations, which hold for all
$F,F_1,F_2\in\DDd,\lambda_1,\lambda_2\in\CC$.
\begin{description}
  \item[Q1.] Linearity:
             $\Xi(\lambda_1 F_1+\lambda_2 F_2)=\lambda_1\Xi(F_1)+\lambda_2\Xi(F_2)$.
  \item[Q2.] Adjoint: $\Xi(\Gamma F)=\Xi(F)^\ast$.
  \item[Q3.] Field equation: $\Xi(DF)=0$.
  \item[Q4.] Canonical anticommutation relations:
    \begin{equation}
      \Xi\left(F_1\right)\Xi\left(F_2\right)+\Xi\left(F_2\right)\Xi\left(F_1\right)
        =-\mi S\left(F_1,F_2\right)\identity.
    \end{equation}
\end{description}
It should be noted that it is the requirement \textbf{Q4} that contains the essentially `quantum' feature of the algebraic structure.

The usual Dirac field and its Dirac adjoint field are obtained as special cases of the above. For any $h\in\DDcosp$ and
$f\in\DDsp$ we define
\begin{equation}
  \Psi(h) := \Xi\left(\left[\begin{array}{c} h \\ 0 \end{array}\right]\right) \quad \text{and}\quad
  \Psi^+(f) := \Xi\left(\left[\begin{array}{c} 0 \\ f \end{array}\right]\right),
\end{equation}
and interpret them as smeared fields. The charge conjugation map $\psi\mapsto\psi^c$ may be used to
define a $\ast$-automorphism $\alpha_c$ of $\Ff(\Mc,\gb)$ by
\begin{equation}
  \alpha_c\Xi\left(\left[
    \begin{array}{c} h \\ f
    \end{array}\right]\right) =
  \Xi\left(\left[
    \begin{array}{c} -f^{c+} \\ h^{c+}
    \end{array}\right]\right),
\end{equation}
In particular, we have $\alpha_c \Psi(h)= \Psi^+(h^{c+})$, $\alpha_c \Psi^+(f)=-\Psi(f^{c+})$, and $\alpha_c\circ\alpha_c = \id$.

The algebra $\Ff(\Mc,\gb)$ can be endowed with a norm, with respect to which its completion is a $C^\ast$-algebra,
namely the CAR algebra. However we will not need this extra structure below. Finally, we remark that $\Ff(\Mc,\gb)$
should not be regarded as the algebra of observables for this theory, owing to the failure of commutativity at
space-like separation. As in \cite{Dimock}, the net of local algebras should be generated by elements of the form
$\Psi^{+}(f)\Psi(h)$.

\subsection{States, two-point functions and the Hadamard condition}\label{subsection:states}

A state in this framework is a linear functional $\omega:\Ff(\Mc,\gb)\rightarrow\CC$ which is positive [$\omega(A^\ast
A)\geq0$ for all $A\in\Ff(\Mc,\gb)$] and normalized [$\omega(\identity)=1$], with $\omega(A)$ interpreted as the
expectation value of observable $A$ in the state $\omega$. A state $\omega$ will be called \emph{charge conjugation
invariant} if it is invariant under $\alpha_c$, so that $\omega(\alpha_c A) = \omega(A)$ for all $A\in\Ff(\Mc,\gb)$. We
will only consider states which are regular enough that the two-point function $\omega_2$, defined by
\begin{equation}
  \omega_2\left(F_1\otimes F_2\right):=\omega\left(\Xi(F_1)\Xi(F_2)\right),
    \quad F_1,F_2\in\DDd,
\end{equation}
is a continuous linear functional on $\DDd\otimes\DDd$. In this case we may introduce distributions $\Wtwo$ and
$\Wtwo^\Gamma$ on $\DDsp\otimes\DDcosp$, defined by\footnote{Comparison with equations (2.49) and (2.50) of
\cite{Fewster/Verch_1} reveals that the two-point functions $\omega_Q$ and $\omega_Q^\Gamma$ defined there are equal to
our $\Wtwo$ and $\Wtwo^\Gamma$ for the special case in which $\omega$ is quasi-free. We emphasize that our treatment is
not restricted to quasi-free states.}
\begin{eqnarray}
  \Wtwo\left(f\otimes h\right) &:=&
    \omega_2\left(\left[\begin{array}{c} 0 \\
      f \end{array}\right]\otimes\left[\begin{array}{c} h \\
      0 \end{array}\right]\right)=
    \omega\left(\Psi^{+}\left(f\right)\Psi\left(h\right)\right), \\
  \Wtwo^\Gamma\left(f\otimes h\right) &:=&
    \omega_2\left(\left[\begin{array}{c} h \\
      0 \end{array}\right]\otimes\left[\begin{array}{c} 0 \\
      f \end{array}\right]\right)=
    \omega\left(\Psi\left(h\right)\Psi^{+}\left(f\right)\right),
\end{eqnarray}
where $f\in\DDsp,h\in\DDcosp$. The distributions $\Wtwo$ and $\Wtwo^\Gamma$ will also be referred to as two-point
functions. Given a reference state $\omega_0$, with corresponding
two-point functions $\Wtwo_0$ and $\Wtwo^\Gamma_0$, we may also define
normal-ordered two-point functions:
\begin{eqnarray}
  \norder{\Wtwo}&=&\Wtwo-\Wzerotwo \\
  \norder{\Wtwo}^\Gamma&=&\Wtwo^\Gamma-\Wzerotwo^\Gamma.
\end{eqnarray}

An important fact, arising as a direct consequence of Hermiticity (axiom \textbf{Q2}) and positivity of states, is that $\omega_2$ is a
distribution of \emph{positive type}, in the sense that
\begin{equation}
  \omega_2\left(\Gamma\left(F\right)\otimes F\right)\geq0\quad\forall F\in\DDd.
\end{equation}
The following positivity properties of $\Wtwo$ and $\Wtwo^\Gamma$ follow immediately:
\begin{eqnarray}
  \Wtwo\left(f\otimes f^{+}\right)&\geq&0\quad\forall f\in\DDsp, \\
  \Wtwo^\Gamma\left(h^{+}\otimes h\right)&\geq&0\quad\forall h\in\DDcosp, \\
  \Wtwo\left(h^{+}\otimes h\right)&\geq&0\quad\forall h\in\DDcosp, \\
  \Wtwo^\Gamma\left(f\otimes f^{+}\right)&\geq&0\quad\forall f\in\DDsp.
\end{eqnarray}
Notice that the anti-commutation relations
\begin{equation}
  \fl
  \Psi^{+}\left(f\right)\Psi\left(h\right)+\Psi\left(h\right)\Psi^{+}\left(f\right)
    =-\mi\langle h,\Ssp f \rangle\identity,
    \quad h\in\DDcosp,f\in\DDsp
\end{equation}
for the field $\Psi$ and its adjoint, when evaluated in a state $\omega$, lead to the result
\begin{equation}
  \Wtwo\left(f\otimes h\right)+\Wtwo^\Gamma\left(f\otimes h\right)
    =-\mi\langle h,\Ssp f \rangle,
    \quad h\in\DDcosp,f\in\DDsp.
\end{equation}
Observe that the right-hand side is independent of the state chosen (the states are normalized, by definition). It then
follows that the normal-ordered two-point functions satisfy
\begin{equation}
  \norder{\Wtwo}\left(f\otimes h\right)=-\norder{\Wtwo}^\Gamma\left(f\otimes h\right),
    \quad h\in\DDcosp,f\in\DDsp.
\end{equation}
It is also easily seen that, if $\omega$ is charge-conjugation invariant, then
\begin{equation}\label{eq:cconinvforW}
  \Wtwo^\Gamma\left(f\otimes h\right) = -\Wtwo\left(h^{c+}\otimes f^{c+}\right).
\end{equation}

For the remainder of this paper we will restrict to the class of Hadamard states, which are distinguished by the
singularity structure of their two-point functions. For scalar fields, the Hadamard condition was originally formulated
in terms of the so-called Hadamard series \cite{Kay/Wald}; however, Radzikowski \cite{Radzikowski} realized that the
condition was equivalent to demanding a particular form for the wave-front set \cite{Hormander} of the two-point
function. As we will use only those features of the wave-front set which have been used before in the context of QEIs,
we refer the reader to \cite{Fewster 2000, Fewster/Verch_1} for the relevant background.

Radzikowski's reformulation of the Hadamard condition was extended to Dirac fields in \cite{Kratzert, Kohler, Hollands} (see also
\cite{Najmi/Ottewill, Sahlmann/Verch} for the Hadamard series in this connection). The upshot is that a (not necessarily charge
conjugation invariant) state $\omega$ on $\Ff(\Mc,\gb)$ is Hadamard if and only if the wave-front set of its two-point function
$\omega_2$ satisfies the \emph{micro-local spectrum condition}
\begin{equation}
  \fl
  \WF\left(\omega_2\right)=\left\{\left(p,\xi;p^\prime,-\xi^\prime\right)
    \in\dot{T}^\ast\left(\Mc\times\Mc\right) \mid
      \left(p,\xi\right)\sim\left(p^\prime,\xi^\prime\right);\quad\xi\in\mathcal{N}^{+}_p
    \right\}.
\end{equation}
Here $\dot{T}^\ast(\Mc\times\Mc)$ is the cotangent bundle---without the zero section---over $\Mc\times\Mc$,
$(p,\xi)\sim(p^\prime,\xi^\prime)$ means that there is a light-like geodesic connecting the points $p$ and $p^\prime$ in
$\Mc$, to which $\xi$ and $\xi^\prime$ are co-tangent, and along which $\xi$ and $\xi^\prime$ are related by parallel
transport. Finally, $\mathcal{N}^{+}_p$ is the set of all future-directed null covectors at $p$.

For Hadamard states, the two-point functions $\Wtwo$ and $\Wtwo^\Gamma$ satisfy the micro-local spectrum conditions
\begin{eqnarray}
  \fl
  \WF\left(\Wtwo\right) & = & \left\{
    \left(p,\xi;p^\prime,-\xi^\prime\right)
    \in\dot{T}^\ast\left(\Mc\times\Mc\right) \mid
      \left(p,\xi\right)\sim\left(p^\prime,\xi^\prime\right);\quad\xi\in\mathcal{N}^{+}_p
      \right\}, \\
  \fl
  \WF\left(\Wtwo^\Gamma\right) & = & \left\{
    \left(p,\xi;p^\prime,-\xi^\prime\right)
    \in\dot{T}^\ast\left(\Mc\times\Mc\right) \mid
      \left(p,\xi\right)\sim\left(p^\prime,\xi^\prime\right);\quad\xi\in\mathcal{N}^{-}_p
      \right\}.
\end{eqnarray}
Except for charge conjugation-invariant states, neither of these is
sufficient on its own to prove that the state is Hadamard \cite{Hollands}.

\subsection{Scalar distributions derived from the two-point function}\label{subsection:Ws}

We now introduce various scalar bidistributions obtained from the two-point functions $W$ and $W^\Gamma$, which are needed in the
statement and proof of Theorem~\ref{thm:main}.

Suppose a local section $E$ of $S(\Mc,\gb)$ is given, defined over some open subset $\Nc$ of $\Mc$. Then we may define spinor fields
$E_A$ on $\Nc$ as described at the end of \S\ref{subsec:Geometry}. Given any $\Lambda\in(\DDsp\otimes\DDcosp)^\prime$, we may define a
matrix of scalar bidistributions $\Lambda_{AB}\in\DD^\prime(\Nc\times\Nc)$ ($A,B=1,\ldots,4$) by
\begin{equation}
  \Lambda_{AB}\left(f,h\right) := \Lambda\left(fE_A\otimes hE_B^+\right),\quad f,h\in\DD(\Nc).
\end{equation}
Applied to $W$ and $W^\Gamma$, we obtain matrices $W_{AB}$ and
$W_{AB}^\Gamma$ which are positive type in the sense that, for example,
\begin{equation}\label{eq:postype1}
  \WtwoAB\left(f^A\otimes\overline{f^B}\right) \geq 0,
  \quad\forall f^A\in\DD\left(\Nc\right),
  \quad A=1,2,3,4,
\end{equation}
where we sum over the repeated indices. In particular, the `traces'
\begin{equation}
  \Ws=\delta^{AB}\WtwoAB \quad \text{and}\quad
  \Ws^\Gamma=\delta^{AB}\WtwoAB^{\Gamma}
\end{equation}
are obviously positive type distributions in $\DD^\prime(\Nc\times\Nc)$ by equation (\ref{eq:postype1}).

Several other properties of $\Ws$ and $\Ws^\Gamma$ will be used below.
First, their wave-front sets are easily seen to be constrained by
\begin{eqnarray}\label{eq:WFofWs}
  \WF\left(\Ws\right) &\subset& \WF\left(W\right)|_{\Nc\times\Nc}, \\
  \WF\left(\Ws^\Gamma\right) &\subset& \WF\left(W^\Gamma\right)|_{\Nc\times\Nc},
\end{eqnarray}
because these distributions are simply sums of products of $W$ and $W^\Gamma$ with smooth local sections\footnote{The
external vector bundle tensor product is defined locally as follows. If $B$ and $B^\prime$ are vector bundles over $M$
and $M^\prime$, with projections $\pi$ and $\pi^\prime$, then $B\boxtimes B^\prime$ is a vector bundle over $M\times
M^\prime$ whose fibre at $(p,p^\prime)$ is $\pi^{-1}(p)\otimes\pi^{\prime-1}(p^\prime)$. This is in contrast to the
tensor product $B\otimes B^\prime$ of bundles $B$ and $B^\prime$ over the same base $M$; in particular, $B\otimes
B^\prime$ is again a bundle over $M$.} of the \emph{outer bundle product} $D\Mc\boxtimes D^\ast\Mc$.

Second, let $U^B{}_A$ be any fixed unitary matrix, and define local spinor fields by $E_A^\prime=U^B{}_AE_B$. Then
\begin{eqnarray}\label{eq:independence}
  \Ws\left(f,h\right) &=& \delta^{AB}W\left(fE_A^\prime\otimes h\left(E_B^\prime\right)^+\right) \\
  \Ws^\Gamma\left(f,h\right) &=& \delta^{AB}W^\Gamma\left(fE_A^\prime\otimes h\left(E_B^\prime\right)^+\right)
\end{eqnarray}
for all $f,h\in\DD(\Nc)$, which demonstrates a modest level of independence of $\Ws$, $\Ws^\Gamma$ from the particular
spinor fields used in the construction. This follows because elements of $\DDsp\otimes\DDcosp$ can be identified with
smooth compactly supported sections of $D\Mc\boxtimes D^\ast\Mc$; in particular, $fE_A\otimes hE_B^+$ corresponds to the
section $(p,p^\prime)\mapsto f(p)h(p^\prime)(E_A\boxtimes E_B^+)(p,p^\prime)$. Since
\begin{equation}
  \delta^{AB}E^\prime_A\boxtimes \left(E^\prime_B\right)^+
    = \delta^{AB}U^C{}_A\overline{U^D{}_B}E_C\boxtimes E_D^+
      = \delta^{CD} E_A\boxtimes E_B^+
\end{equation}
because $\delta^{AB}U^C{}_A\overline{U^D{}_B}=(UU^\dagger)^{CD}=\delta^{CD}$ we obtain (\ref{eq:independence}) as
claimed. Some particular instances of this situation are summarized in the following:
\begin{lemma}\label{lemma:invariance}
  Equation (\ref{eq:independence}) holds in the following cases:
  \begin{enumerate}
    \item $E^\prime_A=[E,b^\prime_A]$ for any orthonormal basis $b^\prime_A$ of $\CC^4$.
    \item\label{item:rot} $E^\prime_A=[R_SE,b_A]$ for a fixed $S\in\Spin_0(1,3)$ such that $\Lambda(S)$ is a pure rotation.
    \item\label{item:ccon} $E^\prime_A=E^c_A$
    \item\label{item:dual} $E^\prime_A=E^{A+}$ where $E^A=[E,b_A^\dagger]^\ast$.
  \end{enumerate}
\end{lemma}
{\noindent\em Proof:} We need only check that $E_A^\prime=U^B{}_AE_B$ for some constant unitary $U^B{}_A$ in each of the cases given. In
case (i) we simply observe that there is a unique unitary such that $b_A^\prime=U^{B}{}_A b_B$. For case (ii) we note that
$[R_SE,b_A]=[E,Sb_A]=S^B{}_A[E,b_B]$ and that $S$ is unitary because $\Lambda(S)$ is a rotation (see the remark after
(\ref{eq:S_gamma0})). For case (iii), we observe that $E_A^c=[E,\gamma_0C^\dagger
\overline{b_A}]=(\gamma_0C^\dagger)^B{}_A[E,\overline{b_B}]$ and use the facts that $\gamma_0 C^\dagger$ is unitary in standard
representations, and that $\overline{b_B}$ is an orthonormal basis of $\CC^4$. Finally, in case (iv), since $E_A^\prime=\gamma_0{}^B_A
E_B$, we use the fact that $\gamma_0$ is unitary in standard representations. $\square$

One particular consequence is that, when $\omega$ is charge-conjugation
invariant,
\begin{eqnarray}\label{eq:cconinvWs}
  \Ws^\Gamma\left(f\otimes h\right) &=& \delta^{AB} W^\Gamma\left(fE_A\otimes h E_B^+\right) \nonumber \\
    &=& \delta^{AB} W\left(-h E_B^{+c+}\otimes f E_A^{c+}\right) \nonumber \\
    &=& \delta^{AB} W\left(h E_B^{c}\otimes f E_A^{c+}\right) \nonumber \\
    &=& \Ws\left(h\otimes f\right)
\end{eqnarray}
where we have used (\ref{eq:cconinvforW}), the spinor identity $v^{+c+}=-v^c$, and Lemma~\ref{lemma:invariance}(\ref{item:ccon}).

Third, if we invoke a reference state $\omega_0$ and form its corresponding distributions $\Ws_0$ and $\Ws_0^\Gamma$, we
may define normal-ordered versions $\norder{\Ws}=\Ws-\Ws_0$ and $\norder{\Ws}^\Gamma=\Ws^\Gamma-\Ws_0^\Gamma$, which are
smooth if both $\omega$ and $\omega_0$ are Hadamard. Note that we have $\norder{\Ws}=-\norder{\Ws}^\Gamma$.

Finally, using Lemma~\ref{lemma:invariance}(\ref{item:dual}) let us note that
\begin{equation}\label{eq:pointsplitcharge}
  \fl \Ws\left(f,h\right)
    = \delta_{AB}\omega\left(\Psi^{+}\left(fE^{A+}\right)\Psi\left(hE^B\right)\right)
      = \omega\left(\Psi^{+}\left(fE^{A}\right)\gamma_0{}^A{}_B\Psi\left(hE^B\right)\right)
\end{equation}
so we may interpret $\Ws$ as a point-split unrenormalized charge density with respect to the tetrad $e_a$. Indeed, the
$\norder{\Ws}(p,p)$ is precisely the normal-ordered charge density
\begin{equation}
  \norder{\Ws}\left(p,p\right)= \langle\norder{\Psi^+e_0\cdot\gamma\Psi}\rangle_\omega\left(p\right)
\end{equation}
in this frame. Here, and below, the $\cdot$ denotes contraction of
tensor indices or the metric-induced inner product as appropriate.

\section{The quantum weak energy inequality}
\subsection{The energy density}\label{Section: energy density}

Let $E$ be any local section of $S(\Mc,\gb)$ and define corresponding
spinor fields $E_A$ and tetrad $e_a$. Then the classical stress-energy tensor
has frame components
\begin{equation}
  T_{ab} = \frac{\mi}{2}\left(\psi^+\gamma_{(a}\nabla{}_{b)}\psi - (\nabla{}_{(a}\psi^+)\gamma{}_{b)}\psi\right).
\end{equation}
In \S3 of \cite{Fewster/Verch_1} it is shown how the normal-ordered energy
density, with respect to the given frame and a fixed choice $\omega_0$
of Hadamard reference state, may be obtained by a
point-splitting prescription as
\begin{equation}
  \langle\norder{\rho}\rangle_\omega\left(p\right) = \left(\LL^{AB}\norder{W_{AB}}\right)\left(p,p\right)
\end{equation}
where
\begin{equation}
  \LL^{AB} = \frac{1}{2}\left\{\left[\identity\otimes\left(\mi e_0\cdot\nabla\right)
        -\left(\mi e_0\cdot\nabla\right)\otimes\identity\right]\delta^{AB}
      +\Theta^{AB}\right\}
\end{equation}
and we have defined
\begin{equation}\label{first_expression_for_spin_coefficients}
  \Theta^{AB}:=\mi\left(\delta^{CB}\sigma_0{}^A{}_C
      -\delta^{AC}\overline{\sigma_0{}^B{}_C}\right)
    =\overline{\Theta^{BA}},
\end{equation}
and
\begin{equation}\label{eq: defn of sigma}
  \sigma_0{}^A{}_B:=-\frac{1}{4}\Gamma^a_{0d}\gamma_a{}^A{}_C\gamma^{dC}{}_B.
\end{equation}

There are two main differences between our approach in this paper and that adopted in \cite{Fewster/Verch_1}: first, we make a
particular choice of section $E$ near the sampling world-line $\gamma$ in which the coefficients $\Theta^{AB}$ vanish on $\gamma$;
second, a cleaner treatment of the remaining terms is used to obtain explicit bounds. In the remainder of this subsection we will
describe the first of these elements, leaving the second to the next subsection.

Accordingly, let $\gamma:I\to\Mc$ be a fixed smooth, future-directed time-like curve, parameterized by proper time in an
open interval $I$ of $\RR$ (including the possibility $I=\RR$) and denote its velocity by $u^\mu$. The first step in the
construction of a suitable section $E$ is to select an arbitrary $\tau_0\in I$, and to choose a tetrad $e_a$ at
$\gamma(\tau_0)$ with $e_0^\mu=u^\mu(\tau_0)$. Next, the tetrad is propagated along $\gamma$ by Fermi--Walker transport,
so that $e_0^\mu|_{\gamma(\tau)}=u^\mu(\tau)$ for all $\tau\in I$. Let $\Nc_\gamma$ be the largest open neighbourhood of
$\gamma$ in which every point is joined to $\gamma$ by a unique space-like geodesic meeting $\gamma$
orthogonally,\footnote{To be precise, $\Nc_\gamma$ is the union of all such open neighbourhoods.} and extend the tetrad
into $\Nc_\gamma$ by parallel transport along such geodesics. Finally, the resulting local section of $F(\Mc,\gb)$ may
be lifted to a local smooth section $E$ of $S(\Mc,\gb)$ using the argument set out in \S3 of \cite{Fewster/Verch_1}.
(There are of course two possible lifts.)

Although the construction involves several arbitrary choices, only
limited freedom is available.
\begin{lemma}\label{lemma:rigidity}
  If $E$ and $E^\prime$ are any two local sections of $S(\Mc,\gb)$ obtained in the above fashion, then there is a
  fixed $S\in\Spin_0(1,3)$, with $\Lambda(S)$ a pure rotation, such that $E^\prime=R_SE$ on $\Nc_\gamma$.
\end{lemma}
{\noindent\em Proof:} At any individual point of $\gamma$, the corresponding tetrads $e_a$ and $e^\prime_a$ differ by a
pure rotation, because $e_0=e^\prime_0$. Since Fermi--Walker transport preserves angles, this must be a fixed rotation.
Moreover, the parallel transport used to propagate the tetrads into the remainder of $\Nc_\gamma$ also preserves angles,
so there is a rigid rotation linking the two frames: $e^\prime = R_\Lambda e$ for some rotation $\Lambda$ in $\SO(1,3)$.
At any given point $p$, therefore, $E^\prime_p=R_{S(p)}E_p$, where $S(p)$ is one of the two possible matrices in
$\Spin_0(1,3)$ with $\Lambda(S)=\Lambda$ [recall that these two possibilities differ by a sign]; continuity then imposes
constancy of $S(p)$ on $\Nc_\gamma$. $\square$

The main consequence of this construction has already been mentioned above:
\begin{lemma}\label{lemma:Theta}
  If $E$ is any local section of $S(\Mc,\gb)$ obtained in the above fashion from the curve $\gamma$, then the
  corresponding coefficients $\Theta^{AB}$ vanish identically on $\gamma$.
\end{lemma}
Before giving the proof, we note that the energy density on $\gamma$ may
now be written in the simpler form
\begin{equation}\label{eq:simpler}
  \fl
  \langle\norder{\rho}\rangle_\omega\left(\gamma\left(\tau\right)\right)
    = \left(
        \frac{1}{2}\left[\identity\otimes
          ie_0\cdot\nabla-ie_0\cdot\nabla\otimes\identity\right]
        \norder{\Ws}\right)
    \left(\gamma\left(\tau\right),\gamma\left(\tau\right)\right),
\end{equation}
where $\norder{\Ws}$ is defined as in \S\ref{subsection:Ws}. Furthermore, this expression is independent of the
particular local section $E$ used, provided it is obtained as described above, by Lemma~\ref{lemma:rigidity} and
Lemma~\ref{lemma:invariance}(\ref{item:rot}). This expression will form the basis of our QWEI proof in the next
subsection.

{\noindent\em Proof of Lemma~\ref{lemma:Theta}:} First note that
\begin{equation}
  \Theta^{AB}\left(\tau\right)=\mi\left(\sigma_0{}^{AB}-\overline{\sigma_0{}^{BA}}\right).
\end{equation}
(We drop the $|_{\gamma\left(\tau\right)}$ notation for the moment, assuming all quantities to be evaluated at a fixed
point on $\gamma$.) Now
\begin{equation}
  \sigma_0{}^{AB}-\overline{\sigma_0^{BA}}=-\frac{1}{4}\Gamma^a_{0d}\chi_a{}^d{}^{AB},
\end{equation}
where we have defined
\begin{eqnarray}
  \chi_a{}^d{}^{AB}&:=&\left[\gamma_a\gamma^d\right]^{AB}-\overline{\left[\gamma_a\gamma^d\right]}^{BA} \nonumber\\
    &=&\left[\gamma_a\gamma^d\right]^{AB}-\left[\left(\gamma_a\gamma^d\right)^{\ast}\right]^{AB} \nonumber\\
    &=&\left[\gamma_a\gamma^d-\left(\gamma_a\gamma^d\right)^\ast\right]^{AB} \nonumber\\
    &=&\left[\gamma_a\gamma^d-\gamma^{d\ast}\gamma_a^\ast\right]^{AB}.
\end{eqnarray}
But in a standard representation, we have $\gamma_0^\ast=\gamma_0$ and $\gamma_k^\ast=-\gamma_k$, from which it follows
(on using the anti-commutation relations for the $\gamma_a$) that
\begin{equation}
  \chi_a{}^d{}^{AB}=\left\{
                      \begin{array}{cc}
                        2\left[\gamma_a\gamma^d\right]^{AB} & a\neq0,d\neq0,d\neq a \\
                        0 & \text{otherwise.}
                      \end{array}
                    \right.
\end{equation}
So we are left with
\begin{equation}\label{what's left of the spin-connection}
  \Theta^{AB}\left(\tau\right)=-\frac{\mi}{2}\sum_{a=1}^3\sum_{d\neq a=1}^3\Gamma^a_{0d}|_{\gamma\left(\tau\right)}
    \left[\gamma_a\gamma^d\right]^{AB},
\end{equation}
where we have restored the explicit $|_{\gamma\left(\tau\right)}$ notation.
We remark that the spin connection coefficients $\Theta^{AB}$ do not appear to vanish trivially; secondly,
because the components $\gamma_b{}^A{}_B$ are constant, the variation in $\Theta^{AB}$ along $\gamma$ is
entirely contained in the Christoffel symbols $\Gamma^a_{0d}|_{\gamma\left(\tau\right)}$.

{}From (\ref{defn_of_C_symbols}), we have
\begin{equation}
  \left(e_0\cdot\nabla\right) k=\left(\nabla_0 k^c\right) e_c=\left(\partial_0 k^a+\Gamma^a_{0d}k^d\right) e_a,
\end{equation}
where we have made use of the orthonormality property
$e^a(e_b)=\delta^a_b$. Taking $k=\delta^f_d e_f=e_d$, we have
\begin{equation}\label{use_this_shortly}
  \left(e_0\cdot\nabla\right) e_d=\left(\nabla_0 \delta^c_d\right) e_c
    = \left(\partial_0 \delta_d^a+\Gamma^a_{0d}\right) e_a=\Gamma^a_{0d} e_a.
\end{equation}

Now, the tetrad $\{e_a\}$ is Fermi--Walker transported along $\gamma$, so (substituting $\dot{\gamma}=e_0$ in the
definition (\ref{en: defn of F-W derivative}))
\begin{equation}
  \fl
  \frac{\mathrm{D}_{\text{F-W}}e_d}{\mathrm{D}\tau}
   =
    \left(e_0\cdot\nabla\right) e_d
    -\left[\left(e_0\cdot\nabla\right) e_0\right]\left(e_d\cdot e_0\right)
    +e_0 \left\{e_d\cdot\left[\left(e_0\cdot\nabla\right) e_0\right]\right\}
    = 0.
\end{equation}
Using (\ref{use_this_shortly}), this gives
\begin{equation}
    \Gamma^a_{0d} e_a
    -\Gamma^a_{00} e_a \left(e_d\cdot e_0\right)
    +\Gamma^a_{00} e_0 \left(e_d\cdot e_a\right)
    = 0,
\end{equation}
which, after operating on $e^b$, using $e_a(e^b)=\delta_a^b$, and swapping the indices $a$ and $b$, gives
\begin{equation}
    \Gamma^a_{0d}=
    \Gamma^a_{00} \left(e_d\cdot e_0\right)
    -\Gamma^b_{00} \delta_0^a \left(e_d\cdot e_b\right)
    .
\end{equation}
Consideration of this result reveals that $\Gamma^a_{0d}$ vanishes unless one and only one of $a,d$ is zero. Returning
to equation (\ref{what's left of the spin-connection}) we now see that, if the frame $\{e_a\}$ is Fermi--Walker
transported along $\gamma$, then
\begin{equation}
  \Theta^{AB}\left(\tau\right)= 0.
\end{equation}
This proves the required result.
$\square$

\subsection{Proof of the QWEI}

We briefly summarize the situation at this point. Let $\omega$ and $\omega_0$ be Hadamard states of the Dirac field.
Given any smooth, future-directed curve $\gamma:I\to\Mc$, parameterized by proper time, we have constructed a class of
local sections of $S(\Mc,\gb)$ in an open neighbourhood $\Nc_\gamma$ of $\gamma$ and used this to define distributions
$\Ws$, $\Ws^\Gamma$ (respectively, $\Ws_0$, $\Ws_0^\Gamma$) from the two point-functions of $\omega$ (respectively,
$\omega_0$). These distributions in fact depend only on the states and the curve, rather than the particular section
used (from our class). Furthermore, the energy density along $\gamma$ is given by (\ref{eq:simpler}) in terms of the
smooth normal-ordered quantity $\norder{\Ws}$.

Now fix any real-valued $g\in\Cinfzero(I)$, and choose $\eta\in\Cinfzero(\Nc_\gamma)$
so that $\eta=1$ on a neighbourhood of $\gamma(\supp{g})$. Then
$\eta\otimes\eta\norder{\Ws}$ is smooth and compactly supported in
$\Nc_\gamma\times\Nc_\gamma$ and its pull-back
\begin{equation}
  \left(\gamma_2^\ast \eta\otimes\eta\norder{\Ws}\right)\left(\tau,\tau^\prime\right) =
  \left(\eta\otimes\eta\norder{\Ws}\right)\left(\gamma\left(\tau\right),\gamma\left(\tau^\prime\right)\right)
\end{equation}
is smooth and compactly supported in $I\times I$. We extend this to
$\RR\times\RR$ so that it vanishes identically outside $I\times I$.
Using an argument taken from \cite{Fewster/Verch_1},
the smeared energy density
\begin{equation}
  \mathcal{I}:=\int\difx{\tau}\langle\norder{\rho}\rangle_{\omega}\left(\tau\right)g\left(\tau\right)^2
\end{equation}
may now be expressed in the form
\begin{equation}\label{definition of I_0}
  \mathcal{I}=
    \frac{1}{8\pi^2}\int\difx{\lambda}\difx{\lambda^\prime}\left(\lambda+\lambda^\prime\right)
      \widehat{g^2}\left(\lambda-\lambda^\prime\right)
      \left[\gamma_2^\ast\left(\eta\otimes\eta\norder{\Ws}\right)\right]\fathat
        \left(-\lambda,\lambda^\prime\right)
\end{equation}
(see equations (4.3) and (4.4) of \cite{Fewster/Verch_1}). Note that the integral in (\ref{definition of I_0}) is absolutely
convergent, because $\norder{\Ws}$ is smooth. The r\^{o}le of $\eta$ is simply to enforce compact support at this stage; it will be
eliminated at a suitable stage in the argument.

The following identity is the key to our derivation of the QWEI, and is
based on results which also played a r\^{o}le in
\cite{Fewster/Verch_1,Fewster/Mistry}:
\begin{lemma}\label{a lemma}
  If $g\in C^\infty_0(\RR)$ is real-valued, and $F\in\mathscr{E}^\prime(\RR^2)$ is smooth then
  \begin{equation}
    \fl
    \frac{1}{\left(2\pi\right)^2}\int\difx{\lambda}\difx{\lambda^\prime}\left(\lambda+\lambda^\prime\right)
    \widehat{g^2}\left(\lambda-\lambda^\prime\right)
    \widehat{F}\left(-\lambda,\lambda^\prime\right)
    = \frac{1}{\pi}\int\difx{\mu}\mu F\left(g\me_{-\mu}\otimes g\me_\mu\right).
  \end{equation}
\end{lemma}
{\noindent\emph{Proof:}} Applying Lemma~6.1 of \cite{Fewster/Verch_1}, we see that
\begin{equation}
  \left(\lambda+\lambda^\prime\right)\widehat{g^2}\left(\lambda-\lambda^\prime\right)
  =\frac{1}{\pi}\int\difx{\mu}\mu\widehat{g}\left(\lambda-\mu\right)
\overline{\widehat{g}\left(\lambda^\prime-\mu\right)}.
\end{equation}
On the other hand,
\begin{eqnarray}
  \fl
  F\left(g\me_{-\mu}\otimes g\me_\mu\right) &=&  \left((g\otimes g)F\right)\fathat\left(-\mu,\mu\right) \nonumber \\
  &=& \frac{1}{\left(2\pi\right)^2} \int\difx{\lambda}\difx{\lambda^\prime}
  \widehat{F}\left(-\lambda,\lambda^\prime\right)
     \widehat{g}\left(\lambda-\mu\right)\overline{\widehat{g}\left(\lambda^\prime-\mu\right)}
\end{eqnarray}
by the convolution theorem and the fact that $\widehat{g}(-u)=\overline{\widehat{g}(u)}$ because $g$ is real-valued. All
that remains is to justify the interchange of integration order between $\mu$ and $\lambda,\lambda^\prime$. For some
constant $C>0$, estimate
\begin{equation}
  \left|\widehat{g}\left(u\right)\right| \leq C / \left(1+u^2\right).
\end{equation}
Then the arithmetic-geometric mean inequality gives
\begin{eqnarray}
  \fl
  \int\difx{\mu}
  \left|\mu\widehat{g}\left(\lambda-\mu\right)\overline{\widehat{g}\left(\lambda^\prime-\mu\right)}\right|
    &\leq&
  \frac{C}{2}\int\difx{\mu} \left[\frac{\left|\mu\right|}{\left(1+\left(\lambda-\mu\right)^2\right)^2} +
    \frac{\left|\mu\right|}{\left(1+\left(\lambda^\prime-\mu\right)^2\right)^2}\right] \nonumber \\
  &\leq&
  \frac{C\pi}{4}\left(\left|\lambda\right|+\left|\lambda^\prime\right|\right) + C \nonumber \\
  &\leq&
  C\left(1+\left|\lambda\right|+\left|\lambda^\prime\right|\right).
\end{eqnarray}
This result, together with the fact that $\widehat{F}(-\lambda,\lambda^\prime)$ is of rapid decay as
$(\lambda,\lambda^\prime)\rightarrow\infty$, completes the proof by a simple application of Fubini's theorem. $\square$

Applying Lemma~\ref{a lemma} to the averaged energy density $\mathcal{I}$, we have
\begin{equation}
  \mathcal{I}=\frac{1}{2\pi}\int\difx{\mu}\mu
    \gamma_2^\ast\left(\eta\otimes\eta\norder{\Ws}\right)
    \left(g_{\mu}\otimes\overline{g_{\mu}}\right),
\end{equation}
where we have defined $g_\mu:=g\me_{-\mu}$. Because $\eta\circ\gamma$ equals unity on the support of $g$, we may
now discard $\eta$ and write
\begin{eqnarray}
\fl  \mathcal{I} &=& \frac{1}{2\pi}\int_{-\infty}^\infty\difx{\mu}\mu \gamma_2^\ast
    \norder{\Ws}
    \left(g_{\mu}\otimes\overline{g_{\mu}}\right)
      \nonumber\\
\fl      &=&\frac{1}{2\pi}\int_0^\infty\difx{\mu}\mu \gamma_2^\ast
        \norder{\Ws}
        \left(g_{\mu}\otimes\overline{g_{\mu}}\right) +\frac{1}{2\pi}\int_{-\infty}^0\difx{\mu}\mu \gamma_2^\ast
        \norder{\Ws}
        \left(g_{\mu}\otimes\overline{g_{\mu}}\right) \nonumber\\
\fl      &=&\frac{1}{2\pi}\int_0^\infty\difx{\mu}\mu \gamma_2^\ast
        \norder{\Ws}
        \left(g_{\mu}\otimes\overline{g_{\mu}}\right)-\frac{1}{2\pi}\int_{-\infty}^0\difx{\mu}\mu \gamma_2^\ast
        \norder{\Ws}^\Gamma
        \left(g_{\mu}\otimes\overline{g_{\mu}}\right),
\label{eq:main_calc}
\end{eqnarray}
where we have made use of the result $\norder{\Ws}=-\norder{\Ws}^\Gamma$ in the last
step.

So far, we have worked with smooth functions, such as $\norder{\Ws}$ and $\norder{\Ws}^\Gamma$, for which the existence and smoothness
of pull-backs is trivial. We now wish to separate $\norder{\Ws}$ (respectively, $\norder{\Ws}^\Gamma$) into contributions from $\Ws$
and $\Ws_0$ (respectively, $\Ws^\Gamma$ and $\Ws_0^\Gamma$). Even though these are non-smooth distributions the pull-backs exist as
distributions in $\DD^\prime(I\times I)$ by standard techniques in microlocal analysis: the argument is exactly as in \cite{Fewster
2000,Fewster/Verch_1} and will not be repeated here, except to mention that the key issue is that---as can be seen from
(\ref{eq:WFofWs})---their wave-front sets involve only null covectors, which cannot annihilate the time-like tangent vectors of
$\gamma$; furthermore, the wave-front sets of $\gamma_2^\ast\Ws$ and $\gamma_2^\ast\Ws^\Gamma$ are
\begin{eqnarray}
  \WF\left(\gamma_2^\ast\Ws\right) &\subset &
    I\times\RR^{+}\times I\times\RR^{-},\quad\text{and} \\
  \WF\left(\gamma_2^\ast\Ws^\Gamma\right) &\subset &
    I\times\RR^{-}\times I\times\RR^{+}.
\end{eqnarray}
(The same holds, of course, for $\gamma_2^\ast\Ws_0$ and $\gamma_2^\ast\Ws_0^\Gamma$.) Furthermore, the pull-backs inherit the
positive-type property by Theorem~2.2 of \cite{Fewster 2000}. Consequently, on substituting $\gamma_2^\ast\norder{\Ws}=\gamma_2^\ast
\Ws-\gamma_2^\ast \Ws_0$ and $\gamma_2^\ast\norder{\Ws}^\Gamma=\gamma_2^\ast \Ws^\Gamma-\gamma_2^\ast\Ws_0^\Gamma$ in
(\ref{eq:main_calc}) we see that the contribution from the state $\omega$ is nonnegative (note that the second integral runs over
negative values of $\mu$). Discarding this contribution, we obtain the inequality
\begin{equation}
  \fl
  \mathcal{I} \geq
     -\frac{1}{2\pi}\int_0^\infty\difx{\mu}\mu \gamma_2^\ast
      \Ws_0\left(g_{\mu}\otimes\overline{g_{\mu}}\right) + \frac{1}{2\pi}\int_{-\infty}^0\difx{\mu}\mu \gamma_2^\ast
      \Ws_0^\Gamma\left(g_{\mu}\otimes\overline{g_{\mu}}\right)
\end{equation}
or, equivalently,
\begin{equation}
  \mathcal{I}\geq-\frac{1}{2\pi}\int_0^{\infty}\difx{\mu}\mu\left\{
    \gamma_2^\ast\Ws_0\left(g_{\mu}\otimes\overline{g_{\mu}}\right)
   +\gamma_2^\ast\Ws^\Gamma_0\left(g_{-\mu}\otimes\overline{g_{-\mu}}\right)
  \right\}.
\end{equation}
The right-hand side depends only upon the choice of the reference state $\omega_0$ and the curve $\gamma$, as well as the function $g$.
Furthermore, it is manifestly negative. Most importantly, it is finite, because the structure of the wave-front sets of $\gamma_2^\ast
\Ws_0$ and $\gamma_2^\ast \Ws_0^\Gamma$ ensures that the integrand is of rapid decay in $\mu$ as $\mu\rightarrow +\infty$. (Compare, for
example, with the proof of Theorem~4.1 in \cite{Fewster 2000}.)

We therefore have
\begin{equation}
    \int\difx{\tau}\langle\norder{\rho}\rangle_\omega\left(\tau\right)g\left(\tau\right)^2
      \geq
    -\frac{1}{2\pi}\int_0^{\infty}\difx{\mu}\mu \left( S_\mu +S^\Gamma_\mu\right) > -\infty,
\end{equation}
where $S_\mu$ and $S^\Gamma_\mu$ are given by equation (\ref{eq: defn of S mu}). We have thus proved our main result,
Theorem~\ref{thm:main}.

When the reference state is charge conjugation invariant, there is a further simplification, due to the relation (\ref{eq:cconinvWs}).
By arguments similar to those used to prove Theorem~2.2 in \cite{Fewster 2000} the pull-backs have the same
relationship, so
$\gamma_2^\ast\Ws_0^\Gamma(f\otimes h)=\gamma_2^\ast\Ws_0(h\otimes f)$ for $f,h\in\DD(I)$. Consequently, $S^\Gamma_\mu=S_\mu$, and we
have
\begin{equation}
    \int\difx{\tau}\langle\norder{\rho}\rangle_\omega\left(\tau\right)g\left(\tau\right)^2
      \geq
    -\frac{1}{\pi}\int_0^{\infty}\difx{\mu}\mu S_\mu.
\end{equation}

\section{Example: static spacetimes}\label{sec:static}

\subsection{General case}

In this section, we will obtain a simplified form for the bound (\ref{eq: main result}), for the case in which the energy density is
averaged along a static trajectory in a static spacetime, and when the reference state $\omega_0$ is itself static. Further
simplifications occur if $\omega_0$ is charge conjugation invariant, or a ground state. Finally, we show that the simplified bounds
reduce to the bound of \cite{Fewster/Mistry} in Minkowski spacetime, using the Minkowski vacuum (which is, of course, charge conjugation
invariant) as the reference state.

Accordingly, the spacetime $(\Mc,\gb)$ is henceforth assumed to admit a one-parameter group of isometries $\beta_t$,
whose orbits are smooth, time-like and generated by a hypersurface-orthogonal Killing vector field $\xi^\mu$. We choose
$\gamma$ to be one of these orbits and assume, without loss, that $\xi^\mu\xi_\mu=1$ on $\gamma$. In several respects our
argument will resemble that used for scalar fields in \cite{Fewster 2000} (in which stationary spacetimes were treated)
and we will therefore concentrate our attention on those aspects which are different for the Dirac field.

The first task is to promote $\beta_t$ to a one-parameter group of $\ast$-automorphisms of the algebra $\Ff(\Mc,\gb)$;
we employ the strategy outlined in \cite{Strohmaier}. To start, note that $\beta_t$ induces the push-forward
$\bts:T\Mc\rightarrow T\Mc$ and, in an obvious way, an action on the frame bundle $F(\Mc,\gb)$, which we will also
denote $\bts$. This action lifts uniquely to $\bhts:S(\Mc,\gb)\rightarrow S(\Mc,\gb)$ so that
$\widehat{\beta}_{0\ast}=\id_{S(\Mc,\gb)}$, $t\mapsto \bhts E$ is continuous for each $E\in S(\Mc,\gb)$ and $\psi(\bhts
E)=\bts(\psi(E))$, where $\psi:S(\Mc,\gb)\rightarrow F(\Mc,\gb)$ encodes the spin structure. In turn, we may induce
action $\btts$ on $D\Mc$ by $\btts[T,x]_p=[\bhts T,x]_{\beta_t(p)}$, and similarly on $D^\ast\Mc$ by
$\btts[T,x]^\ast_p=[\bhts T,x]^\ast_{\beta_t(p)}$, so that $\btts$ commutes with Dirac adjoint. All these bundle maps
cover the original isometries $\beta_t$, mapping the fibre over $p$ to the fibre over $\beta_t(p)$; they also induce
maps [for which we use the same notation] on the corresponding spaces of (local) sections over these bundles, e.g.,
$(\btts f)(p) = \btts(f(\beta_t^{-1}(p)))$ for $f\in C^\infty(D\Mc)$ etc. The maps $\btts$ act in an obvious way on
$\DDd$, and this action is easily seen to commute with the conjugation $\Gamma$, the doubled Dirac operator $D$ and the
fundamental solutions $S^\pm_{\text{(co)sp}}$, so that $S(\btts F_1,\btts F_2)=S(F_1,F_2)$ for all $F_i\in\DDd$,
$t\in\RR$. In consequence, the map $\alpha_t$ defined by $\alpha_t(\Xi(F)):=\Xi(\btts F)$ extends uniquely to a
unit-preserving $\ast$-automorphism of $\Ff(\Mc,\gb)$.

A state $\omega_0$ is said to be static for $\alpha_t$ if $\omega_0(\alpha_t(A))=\omega_0(A)$ for all
$A\in\Ff(\Mc,\gb)$, $t\in\RR$. In particular, this entails that the corresponding two-point functions obey
\begin{eqnarray}
  \Wzerotwo\left(\btts f\otimes\btts h\right) & = &
    \omega_0\left(\Psi^{+}\left(\btts f\right)\Psi\left(\btts h\right)\right) \nonumber \\
    & = & \omega_0\left(\alpha_t\left(\Psi^{+}\left(f\right)\Psi\left(h\right)\right)\right) \nonumber \\
    & = & \Wzerotwo\left(f\otimes h\right)
\end{eqnarray}
for all $f\in \DDsp$, $h\in\DDcosp$ and similarly,
\begin{equation}
  \Wzerotwo^\Gamma\left(\btts f\otimes\btts h\right)=\Wzerotwo^\Gamma\left(f\otimes h\right).
\end{equation}

We construct a local section $E$ of $S(\Mc,\gb)$ as in \S\ref{Section: energy density}, within the open neighbourhood
$\Nc_\gamma$ of $\gamma$, and claim that it is invariant under the static isometries: $\bhts E=E$. (Clearly,
$\Nc_\gamma$ is also invariant under the Killing flow.) To see this, first note that Fermi--Walker transport and Lie
transport coincide on $\gamma$, because $\xi^\mu$ is hypersurface-orthogonal (see Proposition~\ref{proposition: F-W} of
\ref{Appendix: Fermi-Walker_transport}). Thus the tetrad $e_a$ is Lie transported along $\gamma$. By construction, $e_a$
is propagated into the rest of $\Nc_\gamma$ by parallel propagation along the space-like geodesics meeting $\gamma$
orthogonally. Since (i) the connecting geodesic for a point $\beta_t(q)$ is simply the image under $\beta_t$ of the
connecting geodesic for $q$; (ii) the tangent map $\beta_t^\prime$ intertwines parallel transport along these geodesics,
and (iii) we have already argued that the tetrad is Lie transported along $\gamma$, it follows that
$e_a|_{\beta_t(q)}=\beta_t^\prime e_a|_q$ for all $q\in\Nc_\gamma$. Thus $\bts e=e$, and so any smooth lift $E$ to
$S(\Mc,\gb)$ satisfies $\bhts E=E$, while the corresponding local spinor fields obey $\btts E_A=E_A$.

Accordingly, if $f\in\DD(\Mc)$ then $\btts(fE_A) = (\bts f)E_A$, where $\bts f=f\circ
\beta_t^{-1}$. The upshot is that the distributions $\Ws_0,\Ws_0^\Gamma\in \DD^\prime(\Nc\times
\Nc)$ are positive-type distributions obeying
\begin{eqnarray}
  \Ws_0\left(\bts f\otimes \bts h\right) & = & \Ws_0\left(f\otimes h\right), \\
  \Ws_0^\Gamma\left(\bts f\otimes \bts h\right) & = & \Ws_0^\Gamma\left(f\otimes h\right),
\end{eqnarray}
and with wave-front sets sufficiently well-placed that the pull-backs
$\gamma_2^\ast \Ws_0$ and $\gamma_2^\ast \Ws_0^\Gamma$
exist. Arguing exactly as in the Appendix to \cite{Fewster 2000} there must exist $\mathcal{W}_0$,
$\mathcal{W}_0^{\Gamma}\in\DD^\prime(\RR)$ such that
\begin{eqnarray}
    \gamma_2^\ast \Ws_0\left(f\otimes g\right) & = & \mathcal{W}_0\left(f\star\tilde{g}\right), \\
    \gamma_2^\ast \Ws_0^\Gamma\left(f\otimes g\right) & = & \mathcal{W}_0^{\Gamma}\left(f\star\tilde{g}\right),
\end{eqnarray}
where $\tilde{g}(\tau):=g(-\tau)$ and the $\star$ denotes convolution. These distributions are positive-type in the
sense that $\mathcal{W}_0(\overline{f}\star \tilde{f})\ge 0$, for example, and their wave-front sets are easily seen to
obey
\begin{eqnarray}
  \WF\left(\mathcal{W}_0\right) & \subset & \RR\times \RR^+, \\
  \WF\left(\mathcal{W}_0^\Gamma\right) & \subset & \RR\times \RR^-,
\end{eqnarray}
because, for example, $\mathcal{W}_0=\varphi^\ast\gamma_2^\ast \Ws_0$, where $\varphi:\RR\to\RR^2$ is defined by
$\varphi(\tau):=(\tau,0)$. From here we may employ a variant of the Bochner--Schwartz theorem, Theorem~A.11 in
\cite{Fewster/Verch_2}, to deduce that $\mathcal{W}_0$ (respectively, $\mathcal{W}_0^\Gamma$) is a tempered distribution
whose Fourier transform is a polynomially bounded measure such that $(-\infty,u)$ (respectively, $(u,\infty)$) has
finite measure for any $u\in\RR$. Moreover, arguing again as in \S5 of \cite{Fewster 2000},
\begin{equation}
  S_\mu = \widehat{\mathcal{W}_0}\left(|\widehat{g_\mu}|^2\right) =
  \int\frac{\difx{\zeta}}{2\pi} \widehat{\mathcal{W}_0}\left(\zeta\right)\left|\widehat{g}\left(\zeta+\mu\right)\right|^2
\end{equation}
and
\begin{eqnarray}
  S^\Gamma_\mu =
    \widehat{\mathcal{W}_0^\Gamma}\left(\left|\widehat{g_{-\mu}}\right|^2\right)
  & = &
    \int\frac{\difx{\zeta}}{2\pi}\widehat{\mathcal{W}_0^\Gamma}\left(\zeta\right) \left|\widehat{g}\left(\zeta-\mu\right)\right|^2
      \nonumber \\
  & = &
    \int\frac{\difx{\zeta}}{2\pi}\widehat{\mathcal{W}_0^\Gamma}\left(-\zeta\right) \left|\widehat{g}\left(\zeta+\mu\right)\right|^2
\end{eqnarray}
for any real-valued $g\in\Cinfzero(\RR)$. Writing
\begin{equation}
  \mathcal{T}\left(\zeta\right) =
    \widehat{\mathcal{W}_0^\Gamma}\left(\zeta\right)+\widehat{\mathcal{W}_0^\Gamma}\left(-\zeta\right),
\end{equation}
we may then write
\begin{equation}
  \fl
  \int_0^\infty \difx{\mu}\mu\left(S_\mu+S_\mu^\Gamma\right) =
  \int_0^\infty \difx{\mu}\mu \int \frac{\difx{\zeta}}{2\pi}
    \mathcal{T}\left(\zeta\right)\left|\widehat{g}\left(\zeta+\mu\right)\right|^2 =
  \int\difx{u}\left|\widehat{g}\left(u\right)\right|^2 Q\left(u\right)
\end{equation}
where
\begin{equation}\label{eq: defn of Q}
  Q\left(u\right) := \int_{\left(-\infty,u\right)} \frac{\difx{\zeta}}{2\pi} \left(u-\zeta\right)\mathcal{T}\left(\zeta\right)
\end{equation}
is a positive, polynomially bounded function. Accordingly, we have the simplified form of the QWEI bound:
\begin{equation}\label{eq: main result, static case}
  \int\difx{\tau}\langle:\rho:\rangle_\omega\left(\tau\right)g\left(\tau\right)^2
    \geq
  -\frac{1}{2\pi}\int\difx{u}\left|\widehat{g}\left(u\right)\right|^2 Q\left(u\right).
\end{equation}
If, in addition, $\omega_0$ is charge conjugation invariant then we have
$\mathcal{T}(\zeta)=2\widehat{\mathcal{W}_0}(\zeta)$.

Finally, if $\omega_0$ is a ground state, that is, in its GNS representation $\pi_0$, we have
\begin{equation}
  \pi_0\left(\alpha_t\left(A\right)\right) = \me^{-\mi Ht}\pi_0\left(A\right)\me^{\mi Ht}
\end{equation}
for a positive Hamiltonian $H$, then exactly the same arguments used in the Appendix to \cite{Fewster 2000} may be used
to argue that $\widehat{\mathcal{W}_0}$ is supported in $[0,\infty)$, while $\widehat{\mathcal{W}_0^\Gamma}$ is supported in
$(-\infty, 0]$. Thus $\mathcal{T}$ is supported in $[0,\infty)$ and the integration region in the definition (\ref{eq:
defn of Q}) of $Q$ may be restricted to $[0,u)$.

\subsection{Minkowski spacetime}\label{subsec:Minkowski}

Now consider the case of Minkowski spacetime, with $\omega_0$ the charge conjugation invariant ground state. We take $\gamma$ to be the
world line $(t,\mathbf{x}_0)$ of a static observer. Then the bound has the simplified form of equation (\ref{eq: main result, static
case}), but with $Q(u)$ replaced by
\begin{equation}
  Q_{\text{M}}\left(u\right) :=
    \frac{1}{\pi}\int_{\left[0,u\right)}
    \difx{\zeta} \left(u-\zeta\right)\widehat{\mathcal{W}}_0\left(\zeta\right)
\end{equation}
Now it is a standard result that, in un-smeared notation,
\begin{equation}
  \fl
    \omega_0\left(\Psi_A^+\left(x\right)\Psi^B\left(x^\prime\right)\right)
    = \int\frac{\difnx{3}{\kb}}{\left(2\pi\right)^3\left(2\omega_\kb\right)}
        \left(k^a\gamma_a{}^B{}_A + m\delta^B{}_A\right)
          \me_{k}\left(x^\prime-x\right)
\end{equation}
with the latter expression understood as an oscillatory integral, and where $\Psi^B(f):=\Psi(fE^B)$, $\Psi_A^+(f):=\Psi^+(fE_A)$
are the component fields of $\Psi$ and $\Psi^+$. In addition, $k^a=(\omega_\kb,\kb)$ with $\omega_\kb=\sqrt{\|\kb\|^2+m^2}$. It
then follows from (\ref{eq:pointsplitcharge}), along with $\Tr \gamma_0=0$, $\Tr \gamma_a\gamma_b=4\eta_{ab}$ (which hold in any
representation of the Dirac matrices) that
\begin{equation}
  \Ws_0(x,x^\prime)=
    2\int_{\mathbb{R}^3}\frac{\difnx{3}{k}}{\left(2\pi\right)^3}
      \me_{k}\left(x^\prime-x\right).
\end{equation}
It follows that the pull-back $\mathcal{W}_0=\varphi^\ast\gamma_2^\ast\Ws_0$ may be written as
\begin{eqnarray}
  \mathcal{W}_0(t)
    & = & 2\int_{\mathbb{R}^3}\frac{\difnx{3}{\kb}}{\left(2\pi\right)^3}
            \me_{-\omega_\kb}\left(t\right) \nonumber \\
    & = & \frac{1}{\pi^2}
      \int_{\mathbb{R}^+}\difx{k}k^2
        \me_{-\omega_\kb}\left(t\right),
\end{eqnarray}
where we have used spherical polar coordinates to simplify the integral in the second line. Changing the integration variable to
$\zeta = \omega_\kb$, we have
\begin{equation}
  \mathcal{W}_0(t) =
    \frac{1}{2\pi}
      \int_{\mathbb{R}}\difx{\zeta}
        \left[\frac{2}{\pi}\zeta\sqrt{\zeta^2-m^2}\Theta\left(\zeta-m\right)\right]
        \me_{-\zeta}\left(t\right),
\end{equation}
where $\Theta$ here denotes the Heaviside unit step function. The integral is now clearly a Fourier inversion integral,
so that we immediately have
\begin{equation}
  \widehat{\mathcal{W}}_0\left(\zeta\right) =
    \frac{2}{\pi}
      \zeta\sqrt{\zeta^2-m^2}\Theta\left(\zeta-m\right),
\end{equation}
and hence
\begin{eqnarray}
  \fl
  Q_{\text{M}}\left(u\right)
    & = & \frac{2}{\pi^2}\int_{\left[m,u\right)}
    \difx{\zeta} \left(u-\zeta\right)
      \zeta\sqrt{\zeta^2-m^2} \nonumber \\
    \fl
    & = & \frac{1}{12\pi^2}\left[
    u\left(2u^2-5m^2\right)\sqrt{u^2-m^2}
    +3m^4\ln\left(\frac{u+\sqrt{u^2-m^2}}{m}\right)
  \right] \nonumber \\
  \fl & = & \frac{1}{6\pi^2} u^4  \times Q^{\text{D}}_3\left(u/m\right),
\end{eqnarray}
where the function $Q^{\text{D}}_3$ is defined by equation (1.4) of \cite{Fewster/Mistry}.

So, the bound of equation (\ref{eq: main result, static case}) has the final form
\begin{equation}\label{eq:main result, Minkowski case}
  \int\difx{\tau}\langle:\rho:\rangle_\omega\left(\tau\right)g\left(\tau\right)^2
    \geq
  -\frac{1}{12\pi^3}\int_m^\infty \difx{u}\left|\widehat{g}\left(u\right)\right|^2
      u^4 Q^{\text{D}}_3\left(u/m\right)
\end{equation}
in Minkowski spacetime. This is identical to the bound of \cite{Fewster/Mistry}, demonstrating that the approach used
there is generalized here.

\section{Local Covariance}\label{Section:covariance}

A disadvantage of our QWEI is that it is a \emph{difference} quantum energy inequality: it constrains only the
normal-ordered energy density with respect to a reference state. In general spacetimes one would not have access to the
two-point function of a reference Hadamard state in sufficient detail to be able to compute the bound. However, it has
recently been shown, for scalar fields, how difference QEIs may be combined with local covariance to provide constraints
on the \emph{renormalized} stress-energy tensor in locally Minkowskian spacetimes \cite{Fewster/Pfenning_2} (see also
\cite{Marecki05} for the locally Schwarzschild case). This relies on showing that the QEI in question is locally
covariant. Here, we indicate how the Dirac QWEI derived above can be fitted into the locally covariant framework, and
derive a simple consequence. A number of details will be suppressed. Our account of local covariance for Dirac fields is
based on \cite{Dimock,Verch01}; note that an elegant formulation of local covariance in terms of category theory
underlies both of these references and is developed in full in \cite{Brunetti/Fredenhagen/Verch}. We will not use this
language here, but see \cite{Fewster2006} for an account of QEIs in this setting.

Our interest is in the situation where one globally hyperbolic spacetime $(\Mc,\gb)$ with spin structure encoded by
$S(\Mc,\gb)$ and $\psi:S(\Mc,\gb)\to F(\Mc,\gb)$ can be isometrically embedded in another such spacetime
$(\Mc^\prime,\gb^\prime)$, whose spin structure is encoded by $S(\Mc^\prime,\gb^\prime)$ and
$\psi^\prime:S(\Mc^\prime,\gb^\prime)\to F(\Mc^\prime,\gb^\prime)$. We denote the first spacetime and spin structure
simply by $\Mb$, and the second by $\Mb^\prime$. The embedding is required to be compatible with both causality and the
spin structure. To be precise, the embedding $\Thetab$ is a pair $(\Theta,\vartheta)$ of maps $\Theta:S(\Mc,\gb)\to
S(\Mc^\prime,\gb^\prime)$ and $\vartheta:\Mc\to\Mc^\prime$ such that
\begin{enumerate}
  \item $\vartheta$ is a diffeomorphism of $\Mc$ onto its range $\vartheta(\Mc)$ in $\Mc^\prime$.
  \item $\vartheta$ is an isometry ($\vartheta^{\ast}\gb^\prime=\gb$), and preserves orientation and time orientation.
  \item Any causal curve in $(\Mc^\prime,\gb^\prime)$ with endpoints in $\vartheta(\Mc)$ lies entirely in $\vartheta(\Mc)$.
  \item $\Theta$ covers $\vartheta$, in the sense that $\pi^\prime\circ\Theta=\vartheta\circ\pi$, where
    $\pi$ and $\pi^\prime$ are the base-space projections of $S(\Mc,\gb)$ and $S(\Mc^\prime,\gb^\prime)$.
  \item $\Theta$ intertwines the right-actions $R_S$ and $R_S^\prime$ of $\Spin_0(1,3)$ on $S(\Mc,\gb)$ and
    $S(\Mc^\prime,\gb^\prime)$:
    \begin{equation}
      R_S^\prime\circ\Theta = \Theta\circ R_S.
    \end{equation}
  \item $\Theta$ and $\vartheta$ intertwine the spin structures in the sense that
    \begin{equation}
      \psi^\prime\circ\Theta = D\vartheta\circ \psi,
    \end{equation}
    where $D\vartheta$ is the tangent mapping of $\vartheta$.
\end{enumerate}
Under these circumstances, we write $\Thetab:\Mb\to\Mb^\prime$ and say that $\Thetab$ is an \emph{admissible embedding}.
In addition, we may extend the action of $\Theta$ to the spinor and cospinor bundles, defining, for example,
$\widehat{\Theta}:D\Mc\to D\Mc^\prime$ by
\begin{equation}
  \widehat{\Theta}\left[E_p,x\right]_p = \left[\Theta E_p,x\right]_{\vartheta(p)};
\end{equation}
we also use $\widehat{\Theta}$ for the corresponding action on $D^{\ast}\Mc$. These maps induce push-forwards
$\widehat{\Theta}_\ast$ between the smooth compactly supported sections of these bundles and hence between the spaces of
doubled (co)spinors, $\DDdM$ and $\DDdMp$. Further, the map $\Xi_\Mb(F)\mapsto \Xi_{\Mb^\prime}(\widehat{\Theta}_\ast
F)$ may be extended to an injective, unit-preserving $\ast$-homomorphism $\alpha_\Thetab:\Ff(\Mc,\gb)\to
\Ff(\Mc^\prime,\gb^\prime)$. The dual map $\alpha_\Thetab^\ast$ sends Hadamard states on $\Ff(\Mc^\prime,\gb^\prime)$ to
Hadamard states on $\Ff(\Mc,\gb)$.

We may now explain how our Dirac QWEI may be considered as a locally covariant difference QEI. First, on any $\Mb$ we
may form a class $\Tf_\Mb$ of all distributional tensors $\fs$, acting on second rank covariant tensors $\ts$ by
\begin{equation}\label{eq:fdef}
  \fs\left(\ts\right) = \int_I \difx{\tau} g\left(\tau\right)^2u^\mu u^\nu t_{\mu\nu}|_{\gamma\left(\tau\right)}
\end{equation}
where $\gamma:I\to \Mc$ is a time-like curve meeting our usual hypotheses, $u^\mu$ is its velocity and $g$ belongs to
$\Ctinfzero(I;\RR)$, the class of smooth functions with compact connected support contained in $I$ and having no zeros
of infinite order in the interior of their support.\footnote{Our class $\Tf_\Mb$ of sampling tensors was denoted
$\Ff_{\boldsymbol{M}}^{\text{weak}}$ in \cite{Fewster/Pfenning_2}.} The restriction to $\Ctinfzero(I;\RR)$ is fairly
mild, as it is dense in $\Cinfzero(I;\RR)$ \cite{Fewster/Pfenning_2}. Next, define $\Qc_\Mb(\fs,\omega_0)$ for each
$\fs\in\Tf_\Mb$, and Hadamard state $\omega_0$ on $\Ff(\Mc,\gb)$ to equal the right-hand side of (\ref{eq: main
result}), where $\fs$ is related to $I$, $\gamma$ and $g$ by (\ref{eq:fdef}) [note that $I$, $\gamma$ and $g$ may be
reconstructed from $\fs$ up to trivial reparameterizations]. Then in any spacetime, our QWEI takes the form
\begin{equation}
  \langle \Ts_\Mb\left(\fs\right)\rangle_\omega - \langle
    \Ts_\Mb\left(\fs\right)\rangle_{\omega_0} \ge -\Qc_\Mb\left(\fs,\omega_0\right)
\end{equation}
for all Hadamard states $\omega,\omega_0$ on $\Ff(\Mc,\gb)$ and $\fs\in\Tf_\Mb$. Here $\Ts_\Mb$ denotes the renormalized
stress-energy tensor on $\Mb$, so the difference on the left-hand side is precisely the stress-energy tensor
normal-ordered with respect to $\omega_0$. This is the general form of a difference QEI given in
\cite{Fewster/Pfenning_2}; what remains is to verify that it is locally covariant.

To establish covariance, we must show that, given any admissible embedding $\Thetab:\Mb\to\Mb^\prime$, we have
$\vartheta_\ast\Tf_\Mb\subset \Tf_{\Mb^\prime}$ and
\begin{equation}\label{eq:Qcovariance}
  \Qc_\Mb\left(\fs,\alpha_\Thetab^\ast\omega^\prime_0\right) = \Qc_\Mb\left(\vartheta_\ast\fs,\omega^\prime_0\right)
\end{equation}
for all $\fs\in\Tf_\Mb$ and Hadamard states $\omega_0^\prime$ on $\Ff(\Mc^\prime,\gb^\prime)$. The first of these
requirements was established in \cite{Fewster/Pfenning_2}. To prove the second, write
$\gamma^\prime=\vartheta\circ\gamma$, $\omega_0 = \alpha_\Thetab^\ast\omega^\prime_0$ and construct distributions
$\Ws_0,\Ws_0^\Gamma\in \DD^\prime(\Nc_\gamma\times\Nc_\gamma)$, and $\Ws^\prime_0,\Ws_0^{\prime\Gamma}\in
\DD^\prime(\Nc_{\gamma^\prime}\times\Nc_{\gamma^\prime})$ from $\omega_0$ and $\omega_0^\prime$ respectively, according
to our usual construction, based on local sections $E$ and $E^\prime$ of $S(\Mc,\gb)$ and $S(\Mc^\prime,\gb^\prime)$.
Below, we will prove:
\begin{lemma}\label{lem:covariance}
  There is an open neighbourhood of $\gamma$ in which $\vartheta^\ast\Ws_0^\prime$ (respectively, $\vartheta_2^\ast\Ws_0^{\prime\Gamma}$)
  coincides with $\Ws_0$ (respectively, $\Ws_0^\Gamma$), where $\vartheta_2:\Mc\times\Mc\to\Mc^\prime\times\Mc^\prime$ is the map
  $\vartheta_2(p,p^\prime)=(\vartheta(p),\vartheta(p^\prime))$.
\end{lemma}
\noindent{}From this, it follows immediately that $\gamma_2^\ast\Ws_0=\gamma_2^{\prime\ast}\Ws_0^\prime$ and
$\gamma_2^\ast\Ws_0^\Gamma=\gamma_2^{\prime\ast}\Ws_0^{\prime\Gamma}$, thus entailing that (\ref{eq:Qcovariance}) holds,
and establishing local covariance.

The following assertion may now be proved on exactly the same lines as Proposition~III.1 of \cite{Fewster/Pfenning_2},
using two facts about the four-dimensional Minkowski space bound obtained in \cite{Fewster/Mistry}: (i) the QWEI bound
for $m>0$ is more stringent than that for $m=0$; (ii) the bound for massless Dirac fields is exactly a factor $4/3$
weaker than its scalar counterpart.
\begin{proposition}
  Let $\Nb$ be a four-dimensional globally hyperbolic
  spacetime with spin structure. Suppose a time-like geodesic segment $\gamma$ of
  proper duration may be enclosed in an admissibly embedded
  subspacetime\footnote{That is, $\Nb^\prime$, endowed with the metric and spin
  structure obtained by restriction from $\Nb$, is admissibly embedded in
  $\Nb$ by the inclusion map.} $\Nb^\prime$ of $\Nb$. If $\Nb^\prime$ may be admissibly embedded in Minkowski space then
  \begin{equation}
    \sup_\gamma \langle \Ts_{\Nb ab}u^a u^b\rangle_\omega \ge
    -\frac{C}{\tau_0^4}
  \end{equation}
  for any Hadamard state $\omega$ of the Dirac field on $\Nb$, where the constant $C=4.226477\dots$
\end{proposition}
\noindent{}The constant $C$ is just $4/3$ of that appearing in Proposition~III.1 of \cite{Fewster/Pfenning_2}. Rather
more stringent bounds are expected for $m>0$ and will be discussed elsewhere. In a similar way, the other results of
\cite{Fewster/Pfenning_2} can be extended to the Dirac case.

\noindent{}It remains to prove Lemma~\ref{lem:covariance}.

\noindent\emph{Proof of Lemma~\ref{lem:covariance}:} Note that
\begin{eqnarray}
  \Ws_0\left(f\otimes h\right) &=&
    \delta^{AB}\alpha_\Thetab^\ast\omega_0^\prime\left(\Psi_\Mb^+\left(fE_A\right)\Psi_\Mb\left(hE_B^+\right)\right) \nonumber \\
  &=&\delta^{AB}\omega_0^\prime\left(\Psi_{\Mb^\prime}^+\left(\widehat{\Theta}_\ast\left(fE_A\right)\right)
    \Psi_{\Mb^\prime}\left(\widehat{\Theta}_\ast(hE_B^+)\right)\right) \nonumber \\
  &=&\delta^{AB}W_0^\prime\left(\widehat{\Theta}_\ast\left(fE_A\right)\otimes\widehat{\Theta}_\ast\left(hE_B^+\right)\right)
\end{eqnarray}
for $f,h\in\DD(\Nc_\gamma)$. By the same arguments as in \S\ref{subsection:Ws}, it is
enough to show that
\begin{equation}\label{eq:enough}
  \delta^{AB}\widehat{\Theta}_\ast E_A\boxtimes\widehat{\Theta}_\ast E_B^+
   = \delta^{AB}E_A^\prime\boxtimes E_B^{\prime+}
\end{equation}
for some open neighbourhood $\mathcal{O}^\prime$ of $\gamma^\prime$, for then
\begin{equation}
  \Ws_0\left(f\otimes h\right) = \Ws_0^\prime\left(\vartheta_\ast f\otimes \vartheta_\ast h\right)
\end{equation}
for all $f,h\in\DD(\mathcal{O})$ where $\mathcal{O}=\vartheta^{-1}(\mathcal{O}^\prime)$. Hence $\Ws_0$ and
$\vartheta_2^\ast\Ws_0^\prime$ coincide on $\mathcal{O}$ and the result is proved.

To establish (\ref{eq:enough}) we observe that
\begin{equation}
  \fl
  \left(\psi^\prime\left(\Theta_\ast E\right)\right)\left(p\right)
    = \psi^\prime\left(\Theta E_{\vartheta^{-1}\left(p\right)}\right)
    = D\vartheta|_{\vartheta^{-1}\left(p\right)}\psi\left(E_{\vartheta^{-1}\left(p\right)}\right)
    = \left(\vartheta_\ast e\right)\left(p\right)
\end{equation}
where $e=\psi(E)$ is the tetrad induced by $E$. Since $\vartheta$ is an isometry, we may deduce that the tetrad
$\vartheta_\ast e$ is Fermi--Walker transported along $\gamma$ and parallel transported along geodesics in
$\mathcal{O}^\prime:=\Nc_{\gamma^\prime}\cap\vartheta(\Nc_\gamma)$ meeting $\gamma^\prime$ orthogonally. Using the
argument of Lemma~\ref{lemma:rigidity}, we must have $\Theta_\ast E=R^\prime_S E^\prime$ on $\mathcal{O}^\prime$ for
some fixed $S\in\Spin_0(1,3)$ with $\Lambda(S)$ a pure rotation. Then (\ref{eq:enough}) follows by
Lemma~\ref{lemma:invariance}(\ref{item:rot}). $\square$

\section{Conclusion}

To conclude, let us compare the Dirac QWEI with the scalar field bound of \cite{Fewster 2000}. The assumptions about the spacetime,
curve $\gamma$, and sampling function $g$ are essentially the same\footnote{The only difference is that here we have permitted curves
parameterized by finite intervals of proper time, but this would require only trivial modifications to \cite{Fewster 2000}.} as those
made here, and the QWEI bound is
\begin{equation}
  \fl
  \int\difx{\tau}\langle\norder{\rho}\rangle_\omega\left(\tau\right)g\left(\tau\right)^2
      \geq
    -\frac{1}{\pi}\int_0^{\infty}\difx{\mu} \left[g\otimes g\gamma_2^\ast\mathcal{T}_0\right]\fathat\left(-\mu,\mu\right) >
  -\infty,
\end{equation}
where $\mathcal{T}_0$ is a distribution defined in $\Nc\times\Nc$ for some open neighbourhood $\Nc$ of $\gamma$, and which is defined
with the aid of a tetrad $e_a$ on $\Nc$, such that $e_0$ coincides with the velocity of the curve on $\gamma$. The freedom introduced by
the choice of tetrad was not explored in \cite{Fewster 2000}; recently, however, it has been noted \cite{Fewster/Pfenning_2} that the
subclass of tetrads which are invariant under Fermi--Walker transport along $\gamma$ all lead to the same value for the bound.

There are therefore several key similarities between the bound presented here, and the scalar bounds (and the spin-$1$ bounds
\cite{Fewster/Pfenning_1}). In particular, the r\^{o}le of Fermi--Walker transport seems worthy of further investigation: do other
choices of tetrad lead necessarily to less stringent bounds within this method? It is also interesting that the Dirac QWEI turns out
to involve the point-split \emph{charge} density; we do not have a good physical understanding as to why this should be.

In terms of applications, we now see that the Dirac QWEI has a simple form in static spacetimes very much along the lines of those for
the scalar and spin-$1$ fields. Thus the Dirac field falls into the abstract QWEI setting that was used in \cite{Fewster/Verch_2} to
investigate the links between the microlocal spectrum condition, QWEIs and the second law of thermodynamics (in the guise of
`passivity'). In fact the Dirac field would be technically easier to analyze and one might expect to close some of the small technical
gaps left in the scalar field case. In addition, we have seen that local covariance can be invoked in conjunction with the QWEI, just
as in the scalar case \cite{Fewster/Pfenning_2} and can used to obtain \emph{a priori} bounds on energy densities in locally
Minkowskian spacetimes.

Finally, we have only discussed QWEIs for the Dirac field, and it would also be interesting to consider more general
QEIs. In the scalar case, such generalizations are quite straightforward (see, for example, \cite{Fewster/Roman_1});
however, it does not appear to be as easy in the Dirac case.

%
%
%

\begin{appendix}

\section{Fermi--Walker transport}\label{Appendix: Fermi-Walker_transport}

In this section we will give an account---brief, but sufficient for our needs---of Fermi--Walker transport. Relevant
references include \cite{Hawking/Ellis,Bini/Jantzen,Poisson}; we also mention that our original inspiration for
investigating Fermi--Walker transport in this context was \cite{Buchholz/Mund/Summers}, where other applications to
quantum field theory are explored.

Let the time-like curve $\gamma(\tau)$, where the parameter $\tau$ is the proper time, be an integral curve of the
vector field $u^\mu$, such that $u^\mu$ is a unit tangent to the curve:
\begin{equation}
  u^\mu=\frac{\mathrm{d}\gamma^\mu}{\mathrm{d}\tau} = \dot{\gamma}^\mu,
    \quad g_{\mu\nu}u^\mu u^\nu = +1.
\end{equation}
The acceleration of $\gamma$ is
\begin{equation}
  a^\mu := u^\nu \nabla_\nu u^\mu
\end{equation}
and (by definition) satisfies $g_{\mu\nu}a^\mu u^\nu=0$.

The Fermi--Walker derivative of a vector field $X$ along $\gamma$ is defined by
\begin{equation}\label{en: defn of F-W derivative}
  \frac{\mathrm{D}_{\text{F-W}}X}{\mathrm{D}\tau} :=
    \frac{\mathrm{D}X}{\mathrm{D}\tau}
    -\frac{\mathrm{D} \dot{\gamma}}{\mathrm{D}\tau} \left(X\cdot\dot{\gamma}\right)
    +\dot{\gamma} \left(X\cdot\frac{\mathrm{D}\dot{\gamma}}{\mathrm{D}\tau}\right)
    ,
\end{equation}
where $\mathrm{D}X/\mathrm{D}\tau:=(\dot{\gamma}\cdot\nabla)X$. Useful expressions, in terms of components, are
\begin{eqnarray}
  \frac{\mathrm{D}_{\text{F-W}}X^\mu}{\mathrm{D}\tau}
  & = & u^\nu\nabla_\nu X^\mu -g_{\sigma\tau}\left(u^\tau a^\mu -a^\tau u^\mu\right)X^\sigma \nonumber \\
  & = & u^\nu\left\{
      \nabla_\nu X^\mu
    -g_{\sigma\tau}\left(u^\tau \nabla_\nu u^\mu - u^\mu\nabla_\nu u^\tau\right)X^\sigma
    \right\}.
\end{eqnarray}
The definition (\ref{en: defn of F-W derivative}), together with the requirement that the Fermi--Walker derivative be
Leibniz and that it commute with contractions, allows the Fermi--Walker derivative of an arbitrary tensor field to be
uniquely determined.

The vector field $X$ is said to be \emph{Fermi--Walker transported} along $\gamma$ if it satisfies
\begin{equation}
  \frac{\mathrm{D}_{\text{F-W}}X}{\mathrm{D}\tau}=0
\end{equation}
everywhere on $\gamma$. An important property of the Fermi--Walker derivative is that the tangent vector field
$\dot{\gamma}$ is automatically preserved under Fermi--Walker transport. Notice also that, if $\gamma$ is a geodesic,
then the Fermi--Walker derivative reduces to the ordinary absolute derivative $\mathrm{D}/ \mathrm{D}\tau\equiv
(\dot{\gamma}\cdot\nabla)$ along $\gamma$.

Finally, consider the important case in which $\gamma$ is a static trajectory in a static spacetime. Then $\gamma$ is one of the smooth,
time-like orbits generated by a hypersurface-orthogonal Killing vector field $\xi$. In this case, we have the following useful result.
\begin{proposition}\label{proposition: F-W}
Let the vector field $X$ be invariant under the Killing flow, so that it is Lie-transported along $\gamma$ with respect
to $\xi$:
\begin{equation}
  \pounds_{\xi}X|_{\gamma} = 0.
\end{equation}
Then $X$ is Fermi--Walker transported along $\gamma$.
\end{proposition}

{\noindent\emph{Proof:}} Since $\xi^\mu\xi_\mu$ is constant on $\gamma$, we can assume, without loss of generality that $\xi^\mu\xi_\mu=1$; this
simply amounts to the proper-time parameterization on $\gamma$. Writing $f=\xi^\mu\xi_\mu$, the acceleration of $\gamma$ is given by $a^\mu =
-\frac{1}{2}\nabla^\mu f$. Since $\pounds_\xi X = 0$, we have $\xi^\mu\nabla_\mu X^\nu = X^\mu\nabla_\mu\xi^\nu$, and so the Fermi--Walker derivative
may be written as
\begin{equation}
  \frac{\mathrm{D}_{\text{F-W}}}{\mathrm{D}\tau}X_\nu =
  X^\mu\left(\nabla_\mu\xi_\nu + \xi_{\left[\mu\right.}\nabla_{\left. \nu \right]}\ln\left|f\right|\right).
\end{equation}
But a hypersurface-orthogonal Killing vector field $\xi$ with $\xi^\mu\xi_\mu\neq0$ satisfies
\begin{equation}
  \nabla_\mu \xi_\nu = -\xi_{[\mu}\nabla_{\nu]}\ln\left|\xi^\sigma\xi_\sigma\right|,
\end{equation}
(see, for example, \cite{Wald}) and so we have the required result. $\square$

\end{appendix}

%
%
%

\end{document}